%% file: MiddlePolice.tex
\documentclass[sigconf]{acmart}

\usepackage{hyperref}
\usepackage{lipsum}
\usepackage{setspace}
\usepackage{booktabs, tabularx}
\usepackage{array, xspace, subfigure, url, xcolor, amsmath, bm, array,amsfonts, balance,multirow}

\definecolor{darkblue}{rgb}{0,0,0.5}
\hypersetup{
	pdftitle={Fine-Grained Endpoint-Driven In-Network Traffic Control for Proactive DDoS Attack Mitigation},
	pdfauthor={Zhuotao Liu, Hao Jin, Yih-Chun Hu, Michael Bailey}
	pdfkeywords={DDoS Attacks, Endpoint-driven, Traffic control},
	colorlinks=true,
	urlcolor=darkblue,
	linkcolor=darkblue,
	citecolor=darkblue
}

\newcommand{\ie}{\emph{i.e.,}\xspace}
\newcommand{\eg}{\emph{e.g.,}\xspace}
\newcommand{\sys}{MiddlePolice\xspace}
\newcommand{\first}{\textsf{(i)}\xspace}
\newcommand{\second}{\textsf{(ii)}\xspace}
\newcommand{\third}{\textsf{(iii)}\xspace}
\newcommand{\forth}{\textsf{(iv)}\xspace}

\newcommand{\mb}{\textsf{mbox}\xspace}
\newcommand{\mbx}{\textsf{mboxes}\xspace}

\makeatletter
\newif\if@restonecol
\makeatother

\usepackage[linesnumbered,ruled,vlined]{algorithm2e}
\usepackage[noend]{algpseudocode}

\SetCommentSty{mycommfont}

\newcommand{\paraspace}{\vspace{0.03in}}
\newcommand{\parab}[1]{\paraspace\noindent{\bf#1}}

\renewcommand\footnotetextcopyrightpermission[1]{} % removes footnote with conference information in first column

\begin{document}
\title{Fine-Grained Endpoint-Driven In-Network Traffic Control for Proactive DDoS Attack Mitigation}
\titlenote{The initial work is published in ACM CCS 2016, titled \emph{\sys: Toward Enforcing Destination-Defined Policies in the Middle of the Internet} \cite{MiddlePolice}.}
\subtitle{Extended Version for \sys~\cite{MiddlePolice} Published in ACM CCS 2016}
\author{
	Zhuotao Liu
}
\affiliation{%
	\institution{University of Illinois at Urbana-Champaign}
}
\email{zliu48@illinois.edu}

\author{
	Hao Jin
}
\affiliation{%
	\institution{Nanjing University}
}
\email{jinhaonju@gmail.edu}

\author{
	Yih-Chun Hu
}
\affiliation{%
	\institution{University of Illinois at Urbana-Champaign}
}
\email{yihchun@illinois.edu}

\author{
	Michael Bailey
}
\affiliation{%
	\institution{University of Illinois at Urbana-Champaign}
}
\email{mdbailey@illinois.edu}

\maketitle

\input{abs}
\input{intro}

\input{survey}
\input{goal}

\input{sys_overview}
\input{design_detail_mbx}

\input{design_detail_filter}

\input{source_auth}
\input{implementation}
\input{evaluation}

\input{related}
\input{discussion}

\input{conclusion}

\section{Acknowledgments}
We thank the anonymous CCS reviewers and all our interviewed industry people for their valuable feedback. This material is based upon work partially supported by NSF under Contract Nos. CNS-1717313, IIP-1758179, CNS-0953600, CNS-1505790 and CNS-1518741. The views and conclusions contained here are those of the authors and should not be interpreted as necessarily representing the official policies or endorsements, either express or implied, of NSF, the University of Illinois, or the U.S. Government or any of its agencies. 

\balance
\bibliographystyle{acm}
\bibliography{paper}

\end{document}

%% file: abs.tex
\section{Abstract}
\iffalse
Volumetric attacks, which overwhelm the bandwidth of a destination,
are amongst the most common DDoS attacks today. One practical approach
to addressing these attacks is to redirect all destination traffic
(\eg via DNS or BGP) to a third-party, DDoS-protection-as-a-service provider 
(\eg CloudFlare) that is well provisioned and equipped with filtering
mechanisms to remove attack traffic before passing the remaining
benign traffic to the destination. An alternative approach 
is based on the concept of network capabilities, 
whereby source sending rates are determined by receiver consent,
in the form of capabilities enforced by the network.
While both third-party scrubbing services and network capabilities
can be effective at reducing unwanted traffic at an overwhelmed destination, 
DDoS-protection-as-a-service solutions outsource
all of the scheduling decisions (\eg fairness, priority and attack
identification) to the provider, while capability-based solutions
require extensive modifications to existing infrastructure to
operate. In this paper we introduce \sys, which seeks to marry
the deployability of DDoS-protection-as-a-service solutions with the
destination-based control of network capability systems. We show that
by allowing feedback from the destination to the provider,
\sys can effectively enforce destination-chosen traffic control policies, 
while requiring no deployment from unrelated parties. 
\fi

Volumetric attacks, which overwhelm the bandwidth of a destination, are among the most common DDoS attacks today. Despite considerable effort made by both research and industry, our recent interviews with over 100 potential DDoS victims in over 10 industry segments indicate that today's DDoS prevention is far from perfect. On one hand, few academical proposals have ever been deployed in the Internet; on the other hand, solutions offered by existing DDoS prevention vendors are not silver bullet to defend against the entire attack spectrum. Guided by such large-scale study of today's DDoS defense, in this paper, we present \sys, the first \emph{readily deployable} and \emph{proactive} DDoS prevention mechanism. We carefully architect \sys such that it requires no changes from both the Internet core and the network stack of clients, yielding instant deployability in the current Internet architecture. Further, relying on our novel capability feedback mechanism, \sys is able to enforce \emph{destination-driven} traffic control so that it guarantees to deliver victim-desired traffic regardless of the attacker strategies. We implement a prototype of \sys, and demonstrate its feasibility via extensive evaluations in the Internet, hardware testbed and large-scale simulations.

%Besides the above technical contributions, we also present what we learned through our industrial interviews with more than $100$ interviewees from over $10$ industry segments that are vulnerable to DDoS attacks. These profound discussions drive us to think what DDoS prevention \emph{really} means for those who need protection, which may offer useful insight for the community to bridge the gap between academic research and industry practice. 

%% file: intro.tex
\section{Introduction}\label{sec:introduction}
Because the Internet internally does not enforce any traffic control requirements, a number of attacks have been developed to overwhelm Internet end systems. The most significant of these attacks is the volumetric Distributed Denial-of-Service (DDoS) attack, representing over 65\% of all DDoS attacks in 2015~\cite{arbor}. In a volumetric DDoS, many attackers coordinate and send high-rate traffic to a victim, in an attempt to overwhelm the bottleneck links close to the victim. Typical Internet links use RED and drop-tail FIFO queuing disciplines, which provide nearly-equal loss rates to all traffic. Consequently, saturated links impose equal loss rates on attacking and legitimate traffic alike. While legitimate traffic tends to back off to avoid further congestion, attack traffic does not back off, so links saturated by a DDoS attack are effectively closed to legitimate traffic. Recent DDoS attacks include a 620 Gbps attack against Krebs' security blog~\cite{krebs} and a 1 Tbps attack against OVH~\cite{ovh}, a French ISP.

Over the past few decades, both the industry and research have made considerable effort to address this problem. Academics proposed various approaches, ranging from filtering-based approaches~\cite{practicalIPTrace, advancedIPTrace,AITF,pushback,implementPushback,StopIt}, capability-based approaches~\cite{siff, TVA, netfence, MiddlePolice}, overlay-based systems~\cite{phalanx,sos,mayday}, systems based on future Internet architectures~\cite{scion,aip,xia} and other approaches~\cite{speakup,mirage, CDN_on_Demand}. Meanwhile, many cloud security service providers (\eg Akamai, Cloudflare) have played an important role in practical DDoS prevention. These providers massively over-provision data centers for peak attack traffic loads and then share this capacity across many customers as needed. When under DDoS attack, victims use DNS or BGP to redirect their traffic to the provider rather than their own networks. These providers apply their proprietary techniques to scrub traffic, separating malicious from benign, and then re-inject the remaining traffic back into the network of their customers.  

Although these academic and industrial solutions have been proposed for years, the research literature offers surprisingly few real-world studies about their performance and the current status of DDoS prevention. Thus, supported by the NSF I-Corps program, we initiated a large-scale interview with over 100 potential DDoS victims spread across more than ten industry segments to understand \first their opinions about academic proposals, \second their current practice for DDoS prevention and \third their opinions about these common practices. Our study highlights multiple key observations, two of which are the indeployability of most academic proposals and the ``band-aid'' nature of the scrubbing services offered by these cloud security service providers. 

Our study clearly establishes that any advanced DDoS prevention system must meet the following two key challenges. 

\noindent\emph{\first Instant Deployability.} Any practical DDoS prevention proposal must be readily deployable in the current Internet architecture. This is challenging because the current Internet contains over 60,000 independent Autonomous Systems (ASes), with varying levels of technological sophistication and cooperativeness. As a result, prior academic approaches (\eg SIFF~\cite{siff}, TVA~\cite{TVA}, and NetFence~\cite{netfence}) that require secret key management and router upgrades across a large number of ASes face significant deployment hurdles. To address this challenge, we carefully architect our propose system \sys such that it requires no changes from both the Internet core and network stack of clients, yielding instant deployability. 

\noindent\emph{\second Proactive Defense.} As we shall see, the filtering services offered by existing Cloud Security Service Providers (CSSPs) are not a silver bullet as the majority of potential DDoS victims that purchase services from CSSPs still experience problems. The fact that even these CSSPs with decades of DDoS mitigation experience cannot develop perfect filtering techniques probably indicates that it is difficult to win the arms race with attackers simply via reactive filtering. Thus, we argue that an advanced DDoS mitigation should focus on delivering victim-desired traffic even without understanding characteristics of attacks. To this end, \sys is designed to enforce destination-driven traffic control for DDoS mitigation so as to minimize any possible disruption caused by the attacks. 

To summarize, our key contributions are:
\begin{itemize}
	\item A study of today's DDoS defense involving over 100 potential DDoS victims in more than 10 industry segments. 
	\item A systematic analysis of our findings to clarify the design space of advanced DDoS prevention mechanisms. 
	\item The design, implementation and evaluation of \sys, the first system offering readily deployable and proactive DDoS mitigation. 
\end{itemize}

%% file: survey.tex
\section{DDoS Defense Today}
Before discussing our proposed system, we first present our study of current status of real-world DDoS attacks and defense. Supported by the NSF Innovation Corps program under grant IIP-1758179, we interviewed more than $100$ security engineers/administrators from over ten industrial segments, including hosting companies, financial departments, online gaming providers, military contractors, government institutes, medical foundations, and existing DDoS prevention vendors. To the best of our knowledge, in the research community, this is the first comprehensive study of DDoS prevention from the perspective of security experts that are the first-line DDoS defenders. Our analysis highlights following key observations. 

\subsection{Deployment of Academic Proposals}
Over the past decades, the research community have proposed various approaches, see discussion in \S~\ref{sec:related}.
%ranging from filtering-based approaches~\cite{practicalIPTrace, advancedIPTrace,AITF,pushback,implementPushback,StopIt}, capability-based approaches~\cite{siff, TVA, netfence}, overlay-based approaches~\cite{phalanx,sos,mayday} and approaches based on future Internet architectures~\cite{scion,aip,xia}. 
From the research perspective, most of these proposals are provably secure and ensure that a DDoS victim can suppress unwanted traffic. Unfortunately, over the past decades, few research proposals have ever been deployed in the Internet. 

Such lack of real-world deployment is because potential DDoS victims (even large ones such as international financial departments) are unable to enforce the deployment of academic proposals in the Internet. For instance, some proposals require software/hardware upgrades from a large number of geographically distributed ASes. However, due to the lack of business relationship with these remote ASes, these victims are incapable of enforcing any deployment within these ASes. Among all our interviews with security administrators, few of them mentioned that they would consider academic proposals for their practical DDoS prevention. 

\subsection{The Market Giants}
In practice, potential DDoS victims rely on DDoS prevention providers to keep them online. In our study, we interviewed ten market giants in this space, including cloud security-service providers that build massively-provisioned data centers to scrub attack traffic and vendors that sell on-premise hardware equipments designed for DDoS prevention. We make the following key observations. 

\subsubsection{Cloud Security-Service Providers}\label{sec:survey:cssp}
Cloud Security-Service Providers (CSSPs) play an important role for DDoS prevention: over 80 percent of our interviewed potential DDoS victims purchase services from CSSPs. Although their actual products could differ, these CSSPs typically work as follows: deploying geographically distributed \emph{automatic systems} (referred to as sites), terminating customer traffic at these sites, filtering traffic using proprietary rules, and reinjecting the scrubbed traffic back to their customers. CSSPs indeed offer invaluable services to their customers. However, their defense is not a silver bullet. In particular, two caveats are worth mentioning. First, connection termination at CSSPs' sites is privacy invasive for some large organizations, such as government institutes and medical foundations. Second, their ``secret sauce'' filtering technique typically deals with large-yet-obvious attacks (\ie although traffic volumes are huge, these attacks are ``trivial'' to distinguish via static filters, such as port filtering). Quoting one administrator: 

\begin{itemize}
	\item[] \emph{I believe automatic systems managed by vendors should only deal with large things  (attacks). Getting into discussion of customer-specific filtering is a slippery slope.} 
\end{itemize}

In \S \ref{sec:survey:victim}, we discuss how these caveats might affect DDoS victims that use protection services from CSSPs.

\subsubsection{On-premise DDoS Prevention Hardware}
Another type of vendor in DDoS prevention space is hardware manufacturers. Unlike CSSPs that build massive infrastructure, these vendors offer specialized hardware with built-in DDoS prevention capability, such as configurable filtering rules, traffic monitoring and advanced queuing. About 45\% of our interviewed potential DDoS targets deploy on-premise hardware. 

\subsection{Potential DDoS Victims}\label{sec:survey:victim}
We interviewed security engineers/administrators from potential DDoS victims that are fairly large organizations whose online presence is critical for their business. Examples include web hosting companies, online content providers, financial departments, medical foundations, government facilities and so on. We summarize our findings as follows. 

\subsubsection{Mixed Opinions Towards CSSPs}
Over 80\% of all interviewed potential DDoS victims purchase scrubbing services from one of these CSSPs. Their opinions towards these services are mixed. Some security administrators said that these providers do a ``decent'' job to prevent attacks. However, the majority of interviewed administrators claim that they experience attacks even after purchasing these services. Although these attacks are not large enough to knock them offline completely, they still result in various performance issues, including system instability, severe packet losses and even customer churn. Quoting one administrator: 

\begin{itemize}
	\item[] \emph{We do understand their (CSSPs) filters are a band-aid. But these are common practice.} 
\end{itemize}

This observation echoes our discussion in \S \ref{sec:survey:cssp}. Since CSSPs mainly focus on dealing with large-yet-obvious attacks, any attack traffic that bypasses their filters will eventually reach end-customers. Although these attacks are ``small'' by a CSSP's definition, they are large enough to cause problems for those DDoS targets. 

\subsubsection{On-Demand Filtering From ISPs}
About 20\% of our interviewed victims (such as medical and goverment institutes) do not rely on CSSPs. The primary reason is privacy concerns: they cannot afford to allow a third party to have full access to their network connections. As a result, they typically deploy on-premise devices to closely monitor network traffic, and once the traffic volume is above a pre-defined threshold, they work together with their ISPs to filter these offending flows. 

\subsection{Survey Summary}
To recap, our survey covers over 100 potential DDoS victims in more than ten industry segments. The key observations are summarized as follows. \first Since most of the academic proposals incur significant deployment overhead in the  Internet, few of them have ever been deployed in the Internet. \second Current CSSPs that dominate the market mainly care about scrubbing large-yet-obvious attacks based on empirical filtering rules. \third Some potential DDoS victims are attacked continuously even if they purchase services from CSSPs. \forth Some potential victims that cannot afford to allow CSSPs to terminate their network connections due to privacy concerns have to rely on their ISPs to block offending flows. 

%% file: goal.tex
\section{Problem Formulation}\label{sec:goals}
\subsection{Design Space and Goals}
Having understood the status of real-world DDoS defense, when designing \sys, we explicitly achieve two primary goals: being readily deployable in the current Internet and offering proactive mitigation even against sophisticated DDoS attacks. To achieve instant deployability, we carefully architect \sys such that it requires no changes from both the Internet core and the network stack of clients. 
%On one hand, the fact that few academic proposals have ever been deployed indicates that probably our research community have not paid much attention to deployability. Thus, in this paper, , yielding instant deployability. On the other hand, 
Further, the fact that even these large CSSPs with decades of DDoS mitigation experience are unable to offer satisfactory defense indicates that probably it is difficult to win the arms race with attackers by only reactive filters. Rather, we argue that effective DDoS mitigation, to a large extent, is about delivering victim-desired traffic even without understanding the characteristics of attacks. Such \emph{proactive} DDoS mitigation can minimize any potential disruption caused by attacks, regardless of how adversaries may adjust their strategies. In summary, to offer readily deployable and proactive DDoS mitigation, \sys enables the following three key properties. 

\parab{Readily Deployable and Scalable.} 
\sys is designed to be readily deployable in the Internet and sufficiently scalable to handle large scale attacks. \emph{To be readily deployable, a system should only require deployment at the destination, and possibly at related parties on commercial terms.} The end-to-end principle of the Internet, combined with large numbers of end points, is what gives rise to its tremendous utility.Because of the diversity of administrative domains, including end points, edge-ASes, and small transit ASes, ASes have varying levels of technological sophistication and cooperativeness. However, some ASes can be expected to help with deployment; many ISPs already provide some sort of DDoS-protection services~\cite{ddos_isp}, so we can expect that such providers would be willing to deploy a protocol under commercially reasonable terms. We contrast this with prior capability-based work, which requires deployment at a large number of unrelated ASes in the Internet and client network stack modification, that violates the deployability model. 

The goal of being deployable and scalable is the pushing reason that \sys is designed to be deployable in cloud infrastructure without changing the Internet core.

\parab{Destination-Driven Traffic Control Policies.} 
\sys is designed to provide the destination with fine-grained control over the utilization of their network resources. Throughout the paper, we use ``destination'' and 	``victim'' interchangeably. Existing cloud-based DDoS-prevention vendors have not provided such functionality. Many previously proposed capability-based systems are likewise designed to work with a single scheduling policy. For instance, CRAFT~\cite{craft} enforces per-flow fairness, Portcullis~\cite{portcullis} and Mirage~\cite{mirage} enforce per-compute fairness, NetFence~\cite{netfence} enforces per-sender fairness, SIBRA \cite{SIBRA} enforces per-steady-bandwidth fairness, and SpeakUp \cite{speakup} enforces per-outbound-bandwidth fairness. If any of these mechanisms is ever deployed, a single policy will be enforced, forcing the victim to accept the choice made by the defense approach. However, no single fairness regime can satisfy all potential victims' requirements. Ideally, \sys should be able to support arbitrary victim-chosen traffic control policies. In addition to these fairness metrics, \sys can implement ideas such as ARROW's~\cite{one_tunnel} special pass for critical traffic, and prioritized services for premium clients.
 
To the best of our knowledge, \sys is the first DDoS prevention mechanism that advocates destination-driven traffic control for DDoS mitigation. We argue that destination-driven traffic control is the key to offer proactive DDoS attacks even against sophisticated DDoS attacks. There is a fundamental difference between destination-driven traffic control with any vendor or protocol defined traffic control. In particular, compared with vendors or network protocol designers, victim serves can have much more specific and complete knowledge base about their own desired traffic, for instance, through comprehensive traffic monitoring and analysis. As a result, victim servers can make more rational forwarding decisions that are coherent with their mission and applications. Therefore, regardless of how sophisticated a DDoS attack is, as long as \sys can effectively enforce victim-defined traffic control policies to forward victim-preferred traffic, the impact imposed by the DDoS attack on a victim is minimized.

\parab{Fixing the Bypass Vulnerability.} 
Since \sys is designed to be deployed in cloud infrastructure, we need to further address an open vulnerability. In particular, existing cloud-based DDoS-prevention vendors rely on DNS or BGP to redirect the destination's traffic to their infrastructures. However, 
this model opens up the attack of infrastructure bypass. For example, a majority of cloud-protected web servers are subject to IP address exposure~\cite{bypass1, bypass2}. Larger victims that SWIP their IP addresses may be unable to keep their IP addresses secret from a determined adversary. In such cases, the adversary can bypass the cloud infrastructure by routing traffic directly 
to the victims. \sys includes a readily deployable mechanism to address this vulnerability.

However, it is non-trivial to achieve the above key properties simultaneously. Existing research literature is replete with capability-based systems that provide a single fairness guarantee with extensive client modification and deployment at non-affiliated ASes. The novelty and challenge of \sys is therefore architecting a system to move deployment to the cloud while enforcing a wide variety of destination-selectable fairness metrics. Built atop a novel capability feedback mechanism, MiddlePolice meets the challenge,  thereby protecting against DDoS more flexibly and deployably.

\subsection{Adversary Model and Assumptions}\label{subsec:assumptions}
\parab{Adversary Model.} We consider a strong adversary owning large botnets that can launch strategic attacks and amplify its attack~\cite{amplification}. We assume the adversary is not on-path between any \mb and the victim, since otherwise it could drop all packets. Selecting routes without on-path adversaries is an orthogonal problem and is the subject of active research in next-generation Internet protocols (\eg SCION~\cite{scion}).

\parab{Well-connected \mbx.} \sys is built on a distributed and replicable set of \mbx that are well-connected to the Internet backbone. 
We assume the Internet backbone has sufficient capacity and path redundancy to absorb large volumes of traffic, and DDoS attacks against the set of all \mbx can never be successful. This assumption is a standard assumption for cloud-based systems. 

\parab{Victim Cooperation.} \sys's defense requires the victim's cooperation. If the victim can hide its IP addresses from attackers, 
it simply needs to remove a \sys-generated capability carried in each packet and return it back to the \mbx. The victim needs not to modify its layer-7 applications as the capability feedback mechanism is transparent to applications. If attackers can directly send or point traffic (\eg through reflection attacks) to the victim, the victim needs to block the bypassing traffic. \sys includes a packet filtering mechanism 
that is readily deployable on commodity Internet routers with negligible overhead. 

\parab{Cross-traffic Management.} 
We assume that bottlenecks on the path from an \mb to the victim that is shared with other destinations are properly managed, 
such that cross-traffic targeted at another destination cannot cause unbounded losses of the victim's traffic. Generally, per-destination-AS traffic shaping (\eg weighted fair share) on these links will meet this requirement.

%% file: sys_overview.tex
\section{System Overview}\label{sec:sys_overview} 
\begin{figure}[t]
	\centering
	\mbox{
		\includegraphics[width=0.98\columnwidth]{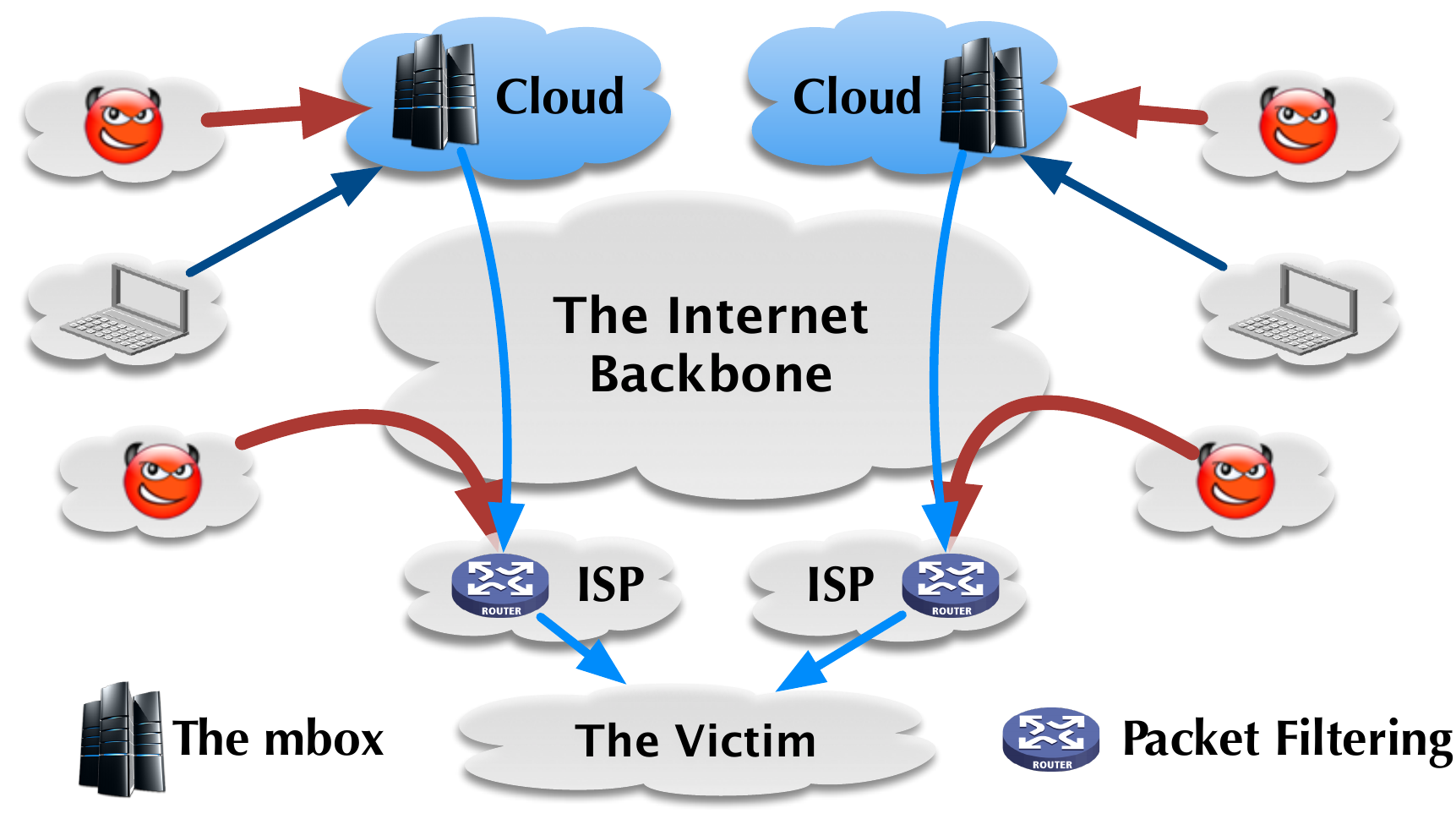}
	}
	\caption{The architecture of \sys. The \mbx police traffic to enforce victim-selected traffic control policies. The packet filtering 
		can discard nearly all traffic bypassing the upstream \mbx. }
	\label{fig:system}
\end{figure}

\sys's high-level architecture is illustrated in Figure~\ref{fig:system}. A \sys-protected victim redirects its traffic to the \mbx. Each \mb polices traversing traffic to enforce the traffic control policy chosen by the victim. The traffic policing relies on a feedback loop of \sys-generated capabilities to eliminate the deployment requirements on downstream paths. When the victim keeps its IP addresses secret, a single deploying \mb can secure the entire downstream path from the \mb to the victim. 

For victims whose IP addresses are exposed, attackers can bypass the \mbx and direct attack traffic to the victim. The same vulnerability applies for current existing cloud-based DDoS-prevention systems. To address this problem, \sys designs a packet filtering mechanism relying on the ACL on commodity routers or switches to eliminate the traffic that does not traverse any \mb.
As long as each bottleneck link is protected by an upstream filter, the bypass attack can be prevented.

%% file: design_detail_mbx.tex
\section{Detailed Design of \mbx}\label{sec:police_algo}
\sys's traffic policing algorithm \first probes the available downstream bandwidth from each \mb to the victim and \second allocates the bandwidth to senders to enforce the traffic control policies chosen by the victim. 

\parab{Bandwidth Probe.} The fundamental challenge of estimating downstream bandwidth is that \sys requires no deployment at downstream links. Such a challenge is two-sided: an overestimate will cause downstream flooding, rendering traffic policing useless, 
while an underestimate will waste downstream capacity, affecting networking performance. 

To solve the overestimation problem, \sys relies on a \emph{capability feedback} mechanism to make senders self-report how many packets they have successfully delivered to the victim. Specifically, upon a packet arrival, the \mb stamps an unforgeable capability in the packet. When the packet is delivered to the victim, \sys's capability handling module (CHM) deployed on the victim returns the carried capability back to the \mb. If the capability is not returned to the \mb after a sufficiently long time interval (compared with the RTT between the \mb and victim), the \mb will consider the packet lost. Thus, the feedback enables the \mb to infer a packet loss rate (hereinafter, LLR) for each sender. Then the \mb estimates the downstream capacity as the difference between the number of packets received from all senders and packets lost on the downstream path. As the estimation is based on the traffic volume \emph{delivered} to the victim, this approach solves the overestimation problem.

However, the above technique does not overcome the underestimation problem. Specifically, since the traffic demand may be less than downstream capacity, simply using the volume of delivered traffic may cause underestimation. To prevent underestimation, 
the \mb categorizes packets from each sender as \emph{privileged packets} and \emph{best-effort} packets. Specifically, the \mb maintains a rate window $\mathcal{W}_R$ for each sender to determine the amount of privileged packets allowed for the sender in each period (hereinafter, \emph{detection period}). $\mathcal{W}_R$ is computed based on the above downstream capacity estimation as well as victim-chosen policies. Packets sent beyond $\mathcal{W}_R$ are classified as best-effort packets. The \mb forwards all privileged packets to the victim, whereas the forwarding decisions for best-effort packets are subject to a short-term packet loss rate (hereinafter, SLR). The SLR reflects downstream packet loss rates (congestion) at a RTT granularity. That is, if the downstream is not congested upon an arrival of a best-effort packet, the \mb will forward the packet. Thus, even when the downstream capacity (and thus $\mathcal{W}_R$) is underestimated, the \mb can still further deliver packets as long as the downstream path is not congested.

\parab{Fairness Regimes.} Each \mb allocates its bandwidth estimate amongst its senders based on the sharing policies chosen by the victim. For policies enforcing global fairness among all senders, all \mbx sharing the same bottleneck share their local observations. We consider co-bottleneck detection in \S~\ref{sec:evaluation:large_scale}. 

\subsection{Information Table}\label{sec:information_table}
\newcommand{\itable}{\textsl{iTable}\xspace}
\newcommand{\ctable}{\textsl{cTable}\xspace}
The basis of \sys's traffic policing is an information table (\itable) maintained by each \mb. Each row of the \itable corresponds to a single sender. The contents of the \itable (and therefore its complexity and cost of maintaining the table) depend on the victim-selected traffic control policy; this section describes \itable elements needed for enforcing per-sender fairness, and \S~\ref{sec:bandwidt_sharing_policy} extends the \itable to other traffic control policies.  In \S~\ref{sec:source_authentication}, we describe multiple mechanisms to filter source spoofing at the \mb, so this section ignores source spoofing. 

\begin{table}[t] 
	\centering
	\caption{Fields of an \itable entry and their sizes (bits).}
	\begin{tabular}{|>{}c|>{}c|>{}c|>{}c|>{}c|>{}c|>{}c|>{}c|>{}c|>{}c|}
		\hline
		$f$&
		$\mathcal{T}_A$&
		$\mathcal{P}_{\text{\it id}}$&
		%$\mathcal{R}_T$&
		%$\mathcal{T}_C$&
		%$\mathcal{S}_T$&
		$\mathcal{N}_R$&
		$\mathcal{N}_D$&
		$\mathcal{W}_R$&
		$\mathcal{W}_V$&
		$\mathcal{L}_R$\\
		\hline
		$64$&
		$32$&
		$16$&
		%$1$&
		%$1$&
		$32$&
		$32$&
		$32$&
		$128$&
		$64$\\
		\hline
	\end{tabular}
	\label{tab:flow_entry}
	\normalsize
\end{table}

Each sender $s_i$ has one row in the \itable, identified by a unique identifier $f$.  The table contents are illustrated in Table~\ref{tab:flow_entry}. Other than $f$, the remaining fields are updated in each \emph{detection period}. The timestamp $\mathcal{T}_A$ records the current detection period. The capability ID $\mathcal{P}_{\text{\it id}}$ is the maximum number of distinct capabilities generated for $s_i$. $\mathcal{N}_R$ stores the number of packets received from $s_i$. $\mathcal{N}_D$ indicates the number of best-effort packets dropped by the \mb. $\mathcal{W}_R$ determines the maximum number of privileged packets allowed for $s_i$. The verification window $\mathcal{W}_V$ is designed to compute $s_i$'s packet loss rate, whereas $\mathcal{L}_R$ stores the LLR for $s_i$. 

\subsection{Capability Computation}\label{sec:capability_computation}
For $s_i$, the \mb generates two types of capabilities: distinct 
capabilities and common capabilities. 
The CHM can use either capability to authenticate 
that the packet has traversed the \mb, 
though only distinct capabilities are used to 
infer downstream packet losses. 

A distinct capability for $s_i$ is computed as follows: 
\begin{equation}\label{eq:cap}
\begin{split}
\mathcal{C}  = &~ \text{\it IP}_{\text{\it MP}} ~||~ \text{\it ts} ~||~ \mathcal{P}_{\text{\it id}} ~||~ f ~||~ \mathcal{T}_A ~|| \\ 
&~ \text{\it MAC}_{K_s}(\text{\it IP}_{\text{\it MP}} ~||~ \text{\it ts} ~||~ \mathcal{P}_{\text{\it id}} ~||~ f ~||~ \mathcal{T}_A),
\end{split}
\end{equation} 
where $\text{\it IP}_{\text{\it MP}}$ is the IP address of the \mb issuing $\mathcal{C}$ and 
$\text{\it ts}$ is the current timestamp (included to mitigate replay attack).
The combination of $\mathcal{P}_{\text{\it id}}||f||\mathcal{T}_A$ ensures 
the uniqueness of $\mathcal{C}$. The MAC is computed based on 
a secret key $K_s$ shared by all \mbx. The MAC is $128$ bits, so the 
entire $\mathcal{C}$ consumes ${\sim}300$ bits. 
A common capability is defined as follows
\begin{equation}\label{eq:common_cap}
\mathcal{C}_{c} = \text{\it IP}_{\text{\it MP}}~||~\text{\it ts}~||~\text{\it MAC}_{K_s}(\text{\it IP}_{\text{\it MP}}~||~\text{\it ts}). 
\end{equation}

The design of capability incorporates a MAC to ensure that attackers without secure keys 
cannot generate valid capabilities, preventing capability abuse. 

%To carry capabilities, rather than defining a new packet header, 
%the \mb appends the capabilities to the end of the data 
%payload of the original packets to avoid 
%compatibility problems at the intermediate routers and switches. 
%Upon packet arrival, the CHM trims the capabilities at the packet 
%footer to deliver the original payload to the victim's applications. 
%See detailed implementation is \S \ref{sec:implementation}. 

\subsection{Traffic Policing Logic}\label{sec:rate_limit_logic}
\subsubsection{Populating the \itable}\label{sec:popFlowTable}
We first describe how to populate the \itable. At time $\text{\it ts}$, the \mb receives the first packet from $s_i$. It creates an entry for $s_i$, with $f$ computed based on $s_i$'s source address, and initializes the remaining fields to zero. It then updates $\mathcal{T}_A$ to $ts$, increases both $\mathcal{N}_R$ and $\mathcal{P}_{\text{\it id}}$ by one to reflect the packet arrival and computes a capability using the updated $\mathcal{P}_{\text{\it id}}$ and $\mathcal{T}_A$. 
%To avoid compatibility problems, the \mb appends the capability  
%to the packet's payload, rather than carrying it in a customized packet header.

New packet arrival from source $s_i$ may trigger the \mb to start new detection period for $s_i$. In particular, upon receiving a packet from $s_i$ with arrival time $t_a {-} \mathcal{T}_A > \mathcal{D}_p$ ($\mathcal{D}_p$ is the length of the detection period), the \mb starts a new detection period for $s_i$ by setting $\mathcal{T}_A = t_a$. The \mb also updates the remaining fields based on the traffic policing algorithm (as described in \S~\ref{sec:rate_limit_algorithm}). The algorithm depends on $s_i$'s LLR and the \mb's SLR, the computation of which is described in the following two sections.

% Placing a table between a \section and the first paragraph breaks LaTeX's
% magic that's designed to keep them together.
\def\Th{\text{\it Th}}
\def\id{\text{\it id}}
\def\cap{\text{\it cap}}
\def\rtt{\text{\it rtt}}
\def\slr{\text{\it slr}}
\def\drop{\text{\it drop}}
\def\lpass{\text{\it lpass}}
\begin{table}[t] 
	\centering
	\caption{System parameters. }
	\small
	\resizebox{\columnwidth}{!}{
	\begin{tabular}{|>{}c|>{}c|>{}c|>{}c|>{}c|>{}c|>{}c|>{}c|>{}c|>{}c|}
		\hline
		Symb.&
		Definition&
		Value\\
		\hline
		$\mathcal{D}_p$&
		The length of the detection period&
		$4s$\\
		\hline
		$\text{\it Th}_{\text{\it cap}}$&
		The upper bound of capability ID&
		$128$\\
		\hline
		$\text{\it Th}_{\text{\it rtt}}$&
		Maximum waiting time for cap. feedback&
		$1$s\\
		\hline
		$\text{\it Th}_{\text{\it slr}}^{\text{\it drop}}$&
		SLR thres. for dropping best-effort pkts&
		$0.05$\\
		\hline
		$\beta$&
		The weight of historical loss rates&
		$0.8$\\
		\hline
		$\text{\it Th}_{\text{\it lpass}}$&
		The threshold for calculating LLR&
		$5$\\
		\hline
		$S_{\text{\it slr}}$&
		The length limit of the \ctable&
		$100$\\
		\hline
	\end{tabular}
}
	\label{tab:para}
	\normalsize
\end{table}

\subsubsection{Inferring the LLR for Source $s_i$}\label{sec:infer_LLR}
%The \mb learns each sender's LLR and the overall SLR via the capability 
%feedback loop. Upon receiving a packet, the victim needs to return the carried 
%capability back to the \mb. Any non-received capabilities after a sufficiently long period of 
%time compared with the RTT from the \mb to the victim are considered to be lost. 
%Based on such a feedback loop, \sys can learn packet losses regardless of 
%the location of the bottleneck links. 
%In this following section, we first describe how to learn $s_i$'s LLR. 

\parab{Capability Generation.} For each packet from $s_i$, the \mb generates a distinct capability for the packet if \first its arrival time
$t_a - \mathcal{T}_A < \mathcal{D}_p - \Th_{\rtt}$, and \second the capability ID $\mathcal{P}_{\id} < \Th_{\cap}$. The first constraint ensures that the \mb allows at least $\Th_{\rtt}$ for each capability to be returned from the CHM.  By setting $\Th_{\rtt}$ well above the RTT from the \mb to the victim, any missing capabilities at the end of the current detection period correspond to lost packets. Table~\ref{tab:para} lists the system parameters including $\Th_{\rtt}$ and their suggested values. \sys's performance with different parameter settings is studied and analyzed in \S~\ref{sec:evaluation:large_scale}. The second constraint $\Th_{\cap}$ bounds the number of distinct capabilities issued for $s_i$ in one detection period, and thus bounds the memory requirement. We set $\Th_{\cap}=128$ to reduce the LLR sampling error while keeping memory overhead low.

Packets violating either of the two constraints, if any, will carry a common capability (Equation~(\ref{eq:common_cap})), which is not
returned by the CHM or used to learn $s_i$'s LLR. However, it can be used as an authenticator that the packet has been accepted by one of the upstream \mbx. 

\parab{Capability Feedback Verification.} Let $K_{\text{\it th}}$ denote the number of distinct capabilities the \mb generates for $s_i$, with capability ID ranging from $[1,K_{\text{\it th}}]$. Each time the \mb receives a returned capability, it checks the capability ID to determine which packet (carrying the received capability) has been received by the CHM. $\mathcal{W}_V$ represents a window with $\Th_{\cap}$ bits. Initially all the bits are set to zero. When a capability with capability ID $i$ is received, the \mb sets the $i$th bit in $\mathcal{W}_V$ to one. At the end of the current detection period, the zero bits in the first $K_{\text{\it th}}$ bits of $\mathcal{W}_V$ 
indicate the losses of the corresponding packets. To avoid feedback reuse, the \mb only accepts feedback issued in the current period. 

\parab{LLR Computation.} LLR in the $k$th detection period is computed at the end of the period, \ie the time when the \mb starts a new detection period for $s_i$. $s_i$'s lost packets may contain downstream losses and best-effort packets dropped by the \mb ($\mathcal{N}_D$). The number of packets that $s_i$ sent to downstream links is $\mathcal{N}_R - \mathcal{N}_D$, and the downstream packet loss rate is $\frac{V_0}{\mathcal{P}_{\id}}$, where $V_0$ is the number of zero bits in the first $\mathcal{P}_{\id}$ bits of $\mathcal{W}_V$. Thus, the estimated number of downstream packet losses is $\mathcal{N}_{\text{\it loss}}^{\text{\it dstream}} = (\mathcal{N}_R - \mathcal{N}_D)\frac{V_0}{\mathcal{P}_{id}}$. Then we have $LLR = (\mathcal{N}_{\text{\it loss}}^{\text{\it dstream}} + \mathcal{N}_D) / \mathcal{N}_{R}$.

\def\LLR{\text{\it LLR}}
\def\SLR{\text{\it SLR}}
\def\loss{\text{\it loss}}
\def\dstream{\text{\it dstream}}
\def\recentLoss{\text{\it recentLoss}}
Our strawman design is subject to statistical bias, and may have negative effects on TCP timeouts. In particular, assume one legitimate TCP source recovers from a timeout and sends one packet to probe the network condition. If the packet is dropped again, the source will enter a longer timeout. However, with the strawman design, the source would incorrectly have a $100\%$ loss rate. Adding a low-pass filter can fix this problem: if $s_i$'s $\mathcal{N}_R$ in the current period is less than a small threshold $\Th_{\lpass}$, the \mb sets its LLR in the current period to zero. Attackers may exploit this design using on-off attacks~\cite{low-rate}. In \S~\ref{sec:rate_limit_algorithm}, we explain how to handle such attacks.  Formally, we write $\LLR$ as follows  

\begin{equation}\label{equ:llr}
\LLR = \begin{cases} 0, & \mbox{if } \mathcal{N}_R < \Th_{\lpass} \\ 
\frac{\mathcal{N}_{\loss}^{\dstream} + \mathcal{N}_{D}}{\mathcal{N}_{R}}, & \mbox{otherwise} \end{cases}
\end{equation}

\subsubsection{Inferring the SLR}\label{sec:infer_SLR}
SLR is designed to reflect downstream congestion on a per-RTT basis (\ie almost realtime downstream congestion), and is computed across all flows from the \mb to the victim. Like LLR, SLR is learned through capabilities returned by the CHM. Specifically, the \mb maintains a hash table (\ctable) to record capabilities used to learn its SLR. The hash key is the capability itself and the value is a single bit value (initialized to zero) to indicate whether the corresponding key (capability) has been returned. 

As described in \S~\ref{sec:infer_LLR}, when a packet arrives (from any source), the \mb stamps a capability on the packet. The capability will be added into \ctable if it is not a common capability and \ctable's length has not reached the predefined threshold 
$\mathcal{S}_{\slr}$. The \mb maintains a timestamp $\mathcal{T}_{\slr}$ when the last capability is added into \ctable. Then, 
it uses the entire batch of capabilities in the \ctable to learn the SLR. We set $\mathcal{S}_{\slr}=100$ to allow fast capability loading to the \ctable (and therefore fast downstream congestion detection), while minimizing sampling error from $\mathcal{S}_{\slr}$ being too small. 

The \mb allows at most $\Th_{\rtt}$ from $\mathcal{T}_{\slr}$ to receive feedback for all capabilities in the \ctable. Some capabilities may be returned before $\mathcal{T}_{\slr}$ (\ie before the \ctable fills). Once a capability in \ctable is returned, the \mb marks it as received. Upon receiving a new packet with arrival time $t_a > \mathcal{T}_{\slr} + \Th_{\rtt}$, the \mb computes $\SLR=\frac{Z_0}{\mathcal{S}_{\slr}}$, where $Z_0$ is the number of \ctable entries that are not received. The \mb then resets the current \ctable to be empty to start a new monitoring cycle for SLR. 

\newcommand{\algrule}[1][.2pt]{\par\vskip.5\baselineskip\hrule height #1\par\vskip.5\baselineskip}
\begin{algorithm}[!t]
\SetKwData{CG}{CapabilityHandling}
\SetKwData{cTH}{cTableHandling}
\SetKwData{iTH}{iTableHandling}
\SetKwData{iTR}{iTableEntryRetrieval}
\SetKwData{BH}{BestEffortHandling}
\SetKwData{FC}{\textbf{Function:}}
\SetKwData{IP}{\textbf{Input:}}
\SetKwData{OP}{\textbf{Output:}}
\SetKwData{BS}{BandwidthAllocationPolicy}
\SetKwData{Main}{\textbf{Main Procedure:}}

\small

\IP \\
\quad Packet $P$ arrived at time $\text{\it ts}$;\\
\OP\\
\quad \itable updates and possible \ctable updates; \\
%\quad Possible updates in \ctable;\\
\quad The forwarding decision of $P$;\\
	
\algrule[0.5pt]

\Main \\
	\Begin{
		$\mathcal{F} \leftarrow$ \iTR{$P$}\; \label{line:entry_retri}
		$\mathcal{F}.\mathcal{N}_R \leftarrow \mathcal{F}.\mathcal{N}_R + 1$\;
			\eIf{$\mathcal{F}.\mathcal{N}_R < \mathcal{F}.\mathcal{W}_R$}{\tcc{Privileged packets} \label{line:prvi_or_bs}
				\CG{$P$, $\mathcal{F}$}\; \label{line:cap_handling_call}
				Append $P$ to the privileged queue\;
			}{ \tcc{Best-effort packets}
				\BH{$P$, $\mathcal{F}$}\;
			}
		
		\tcc{Starting a new detection period if necessary}
		\lIf{$\text{\it ts} - \mathcal{F}.\mathcal{T}_A > \mathcal{D}_p$}{\iTH{$\mathcal{F}$}} \label{line:new_period_call}
	}

\algrule[0.5pt]
\FC \CG{$P$, $\mathcal{F}$}:\label{line:cap_handling}\\ 
\Begin{
  \tcc{Two constraints for distinct-capability generation}
 \eIf{$\mathcal{F}.\mathcal{P}_{\id} < \Th_{\cap}$ \textbf{and} 
 	$ts {-} \mathcal{F}.\mathcal{T}_A < \mathcal{D}_p {-} \Th_{\rtt}$}
	 {  
	 	$\mathcal{F}.\mathcal{P}_{\id} \leftarrow \mathcal{F}.\mathcal{P}_{\id} + 1$\;
	 	Generate capability $\mathcal{C}$ based on Equation (\ref{eq:cap})\;
	 	%Append $\mathcal{C}$ into $P$'s payload\;
	 	\cTH{$\mathcal{C}$}\;
	 }
	 {\tcc{Common capability for packet authentication}
	 	Generate capability $\mathcal{C}_c$ based on Equation (\ref{eq:common_cap})\;
	 	%Append $\mathcal{C}_c$ into $P$'s payload;
	 }
}

\algrule[0.5pt]
\FC \BH{$P$, $\mathcal{F}$}:\label{line:best_effort} \\ 
\Begin{
	\If{SLR  $ < \Th_{\slr}^{\drop}$ \textbf{and}  $\mathcal{F}.\mathcal{L}_R < \Th_{\slr}^{\drop}$}{
		\CG{$P$, $\mathcal{F}$}\; \label{line:best_effort_capbility}
		Append $P$ to the best-effort queue\;
	}
	\Else{
		Drop $P$; ~~~$\mathcal{F}.\mathcal{N}_D \leftarrow \mathcal{F}.\mathcal{N}_D + 1$\;
		}
}

\algrule[0.5pt]
\FC \iTH{$\mathcal{F}$}:\\ \label{line:new_period}
\Begin{
	Compute $\recentLoss$ based on Equation~(\ref{equ:llr})\;
	\tcc{Consider the historical loss rate}
	$\mathcal{F}.\mathcal{L}_R \leftarrow (1-\beta)\cdot \recentLoss + \beta\cdot \mathcal{F}.\mathcal{L}_R$\;
	%\tcc{Cut the allowed rate if LLR is above a threshold}
	%\If{$\mathcal{F}.\mathcal{L}_R > Th_{llr}^{cut}$}{ 
		%$\mathcal{F}.\mathcal{W}_R \leftarrow 
		%\min\{\mathcal{F}.\mathcal{W}_R / 2, \mathcal{N}_R - \mathcal{N}_D - \mathcal{N}_{loss}^{dstream}\}$\; \label{line:window_cut}
	%}
	$\mathcal{W}_R \leftarrow$ \BS{$\mathcal{F}$}\; \label{line:bandwidth_share}
	Reset $\mathcal{W}_V$, $\mathcal{P}_{\id}$, $\mathcal{N}_R$ and $\mathcal{N}_D$ to zero\;
}

\algrule[0.5pt]
\FC \cTH{$\mathcal{C}$}:\\ \label{line:ctable_handling}
\Begin{
	\tcc{One batch of \ctable is not ready}
	\If{\ctable.\textsl{length} $< \mathcal{S}_{\slr}$}{
		Add $\mathcal{C}$ into \ctable\;
		%\tcc{One batch of \ctable is full}
	}
		\lIf{\ctable.\textsl{length} $== \mathcal{S}_{\slr}$}{
			$\mathcal{T}_{\slr} \leftarrow \text{\it ts}$
		}
}
\caption{Traffic Policing Algorithm.}\label{alg:rateLimit}
\end{algorithm}
\normalsize

\subsubsection{Traffic Policing Algorithm}\label{sec:rate_limit_algorithm}
We now detail the traffic policing algorithm to enforce victim-selected traffic control policies. We formalize the traffic policing logic in Algorithm~\ref{alg:rateLimit}. Upon receiving a packet $P$, the \mb retrieves the entry $\mathcal{F}$ in \itable matching $P$ (line~\ref{line:entry_retri}). If no entry matches, the \mb initializes an entry for $P$. 

$P$ is categorized as a privileged or best-effort packet based on $\mathcal{F}$'s $\mathcal{W}_R$ (line~\ref{line:prvi_or_bs}). All privileged packets are accepted, whereas best-effort packets are accepted conditionally. If $P$ is privileged, the \mb performs necessary capability handling (line~\ref{line:cap_handling_call}) before appending $P$ to the privileged queue. The \mb maintains two FIFO queues to serve all \emph{accepted} packets: the privileged queue serving privileged packets and the best-effort queue serving best-effort packets. The privileged queue has strictly higher priority than the best-effort queue at the output port. \textsf{CapabilityHandling} (line~\ref{line:cap_handling}) executes the capability generation and \ctable updates (line~\ref{line:ctable_handling}), as detailed in \S~\ref{sec:infer_LLR} and \S~\ref{sec:infer_SLR}, respectively. 

If $P$ is a best-effort packet, its forwarding decision is subject to the \mb's SLR and $\mathcal{F}'s$ LLR (line~\ref{line:best_effort}). If the SLR exceeds $\Th_{\slr}^{\drop}$, indicating downstream congestion, the \mb discards $P$. Further, if $\mathcal{F}$'s LLR is already above $\Th_{\slr}^{\drop}$, the \mb will not deliver best-effort traffic for $\mathcal{F}$ as well since $\mathcal{F}$ already experiences severe losses. $\Th_{\slr}^{\drop}$ is set to be few times larger than a TCP flow's loss rate in normal network condition~\cite{internetMeasure} to absorb burst losses. If the \mb decides to accept $P$, it performs capability handling (line~\ref{line:best_effort_capbility}). 

Finally, if $P$'s arrival triggers a new detection period for $\mathcal{F}$ (line~\ref{line:new_period_call}), the \mb performs corresponding updates for $\mathcal{F}$ (line~\ref{line:new_period}). To determine $\mathcal{F}$'s LLR, the \mb incorporates both the recent LLR ($\recentLoss$) obtained in the current detection period based on Equation~(\ref{equ:llr}) and $\mathcal{F}$'s 
historical loss rate $\mathcal{L}_R$. This design prevents attackers from hiding their previous packet losses via on-off attacks. 
$\mathcal{F}$'s $\mathcal{W}_R$ is updated based on the victim-selected traffic control policy (line \ref{line:bandwidth_share}), as described below. 

\subsubsection{Traffic Control Policies}\label{sec:bandwidt_sharing_policy} 
We list the following representative bandwidth allocation algorithms that may be implemented in the \textsf{BandwidthAllocationPolicy} module to enforce victim-chosen traffic control policies. We clarify that this list is neither complete nor the ``best''. Instead, these traffic control policies are typical policies used by prior academic papers. Coming up with industry-driven traffic control policies is not the focus of this paper. In fact, our group is conducting active research in this area.

\textsf{NaturalShare}: for each sender, the \mb sets its $\mathcal{W}_R$ for the next period to the number of delivered packets from the sender in the current period. The design rationale for this policy is that the \mb allows a rate that the sender can sustainably transmit without experiencing a large LLR.  

\def\total{\text{\it total}}
\def\size{\text{\it size}}
\textsf{PerSenderFairshare} allows the victim to enforce per-sender fair share at bottlenecks. Each \mb fairly allocates its estimated total downstream bandwidth to the senders that reach the victim through the \mb. To this end, the \mb maintains the total downstream 
bandwidth estimate $\mathcal{N}^{\text{\it total}}_{\text{\it size}}$, which it allocates equally among all senders. To ensure global fairness among all senders, two \mbx sharing the same bottleneck (\ie the two paths connecting the two \mbx with the victim 
both traverse the bottleneck link) share their local observations. We design a co-bottleneck detection mechanism using SLR correlation: if two \mbx' observed SLRs are correlated, they share a bottleneck with high probability. In \S~\ref{sec:evaluation:large_scale}, we evaluate the effectiveness of this mechanism.

\textsf{PerASFairshare} is similar to \textsf{PerSenderFairshare} except that the \mb fairly allocates $\mathcal{N}^{\total}_{\size}$ on a per-AS basis. This policy mimics SIBRA \cite{SIBRA}, preventing bot-infested ASes from taking bandwidth away from legitimate ASes.

\textsf{PerASPerSenderFairshare} is a hierarchical fairness regime: the \mb first allocates $\mathcal{N}^{\total}_{\size}$ on a per-AS basis, and then fairly assigns the share obtained by each AS among the senders of the AS. 

\textsf{PremiumClientSupport} enables the victim to provide premium service to its premium clients, such as bandwidth reservation for upgraded ASes. The victim pre-identifies its premium clients to \sys. \textsf{PremiumClientSupport} can be implemented together with 
the aforementioned allocation algorithms.

%% file: design_detail_filter.tex
\section{Packet Filtering}\label{sec:downstream_authenticator}
When the victim's IP addresses are kept secret, attackers cannot bypass \sys's upstream \mbx to route attack traffic directly to the victim. In this case, the downstream packet filtering  is \emph{unnecessary} since \sys can throttle attacks at the upstream \mbx. However, in case of IP address exposure~\cite{bypass1, bypass2}, the victim needs to deploy a packet filter to discard bypassing traffic. \sys designs a filtering mechanism that extends to commodity routers the port-based filtering of previous work~\cite{bypass_fix, CDN_on_Demand}. Unlike prior work, the filtering can be deployed upstream of the victim as a commercial service.

\subsection{Filtering Primitives}
Although the MAC-incorporated capability can prove that a packet 
indeed traverses an \mb, it requires upgrades from 
deployed commodity routers to perform MAC computation 
to filter bypassing packets. 
Thus, we invent a mechanism based on the existing ACL configurations 
on commodity routers. Specifically, each \mb encapsulates its traversing 
packets into UDP packets (similar techniques have been applied in VXLAN 
and \cite{udp_encap}), and uses the UDP source and destination ports 
(a total of $32$ bits) to carry an authenticator, which is a shared secret 
between the \mb and the filtering point. %Although a $32$-bit 
%secret is not long enough to prevent attackers from generating the secret, 
As a result, a $500$Gbps attack (the largest attack viewed by Arbor Networks~\cite{arbor}) 
that uses random port numbers will be reduced to ${\sim}100$bps since the chance of 
a correct guess is $2^{-32}$. The shared secret can be 
negotiated periodically based on a cryptographically secure 
pseudo-random number generator. 
We do not rely on UDP source address for filtering 
to avoid source spoofing. 
%since attackers may spoof the 
%\mb's IP address. 

\subsection{Packet Filtering Points} 
Deployed filtering points should have sufficient bandwidth so 
that the bypassing attack traffic cannot cause packet losses 
prior to the filtering. 
The filtering mechanism 
should be deployed at, or upstream of, 
each bottleneck link caused by the DDoS attacks. 
For instance, for a victim with high-bandwidth connectivity, if the bottleneck link 
is an internal link inside the victim's network,   
the victim can deploy the filter at the inbound points 
of its network. If the bottleneck link is the link 
connecting the victim with its ISP, the victim can 
can work with its ISP,  on commercially reasonable terms, to deploy 
the filter deeper in the ISP's network 
such that the bypassing traffic cannot 
reach the victim's network. 
Working with the ISPs does not violate
the deployment model in \S \ref{sec:goals} as \sys never 
requires deployment at unrelated ASes.

%% file: source_auth.tex
\section{Source Authentication}\label{sec:source_authentication}
\sys punishes senders even after an offending flow ends. Such persistence can be built on source authentication or any mechanism that maintains sender accountability across flows. In this paper, we propose the following four different source authentication mechanisms, ordered in increasing deployment requirement but increasing robustness.  

\parab{HTTP Redirection-Based Design.} We start with a source authentication that is specific to HTTP/HTTPS traffic. The mechanism ensures that a sender is on-path to its claimed source IP address. This source verifier is completely transparent to clients. The key insight is that the HTTP Host header is in the first few packets of each connection. As a result, the \mb monitors a TCP connection and reads the Host header. If the Host header reflects a generic (not sender-specific) hostname (\eg victim.com), the \mb intercepts this flow, and redirects the connection (HTTP 302) to a Host containing a token cryptographically generated from the sender's claimed source address, \eg $T$.victim.com, where $T$ is the token. If the sender is on-path, it will receive the redirection, and its further connection will use the sender-specific hostname in the Host header. When an \mb  receives a request with a sender-specific Host, it verifies that the Host is proper for the claimed IP source address (if not, the \mb initiates a new redirection), and forwards the request to the victim. Thus, by performing source verification entirely at the \mb, packets from spoofed sources cannot consume any downstream bandwidth from the \mb to the victim. 

If the cloud provider hosting the \mb is trusted (for instance, large CDNs have CAs trusted by all major browsers), the victim can share its key such that HTTPS traffic can be handled in the same way as HTTP traffic. For untrusted cloud  providers~\cite{https_CDN,CDN_on_Demand}, the \mb should relay the encrypted connection to the victim, which performs Host-header-related operations. The victim terminates unverified senders without returning capabilities to the \mb, so the additional traffic from the unverified senders is best-effort under Algorithm~\ref{alg:rateLimit}. In this case, packets from spoofed sources consume limited downstream bandwidth but do not rely on the trustworthiness of the cloud provider. 

We acknowledge that the above mechanism has few weakness. First, it only proves that sender is on path between its claimed IP address to the \mb. Second, if the source validation has to be performed by the victim for HTTPS traffic, it is subject to the DoC attack~\cite{doc} in which attackers flood the victim with new connections to slow down the connection-setup for legitimate clients.

\parab{Multipath-Based Authentication.} We can use the following multipath based mechanism to increase the accuracy of source validation. In particular, since a cloud provider typically has many network points of presence (PoPs), we can reduce the number of potential on-path adversaries by distributing shares of an authentication token to various PoPs with diverse paths back to the sender. To achieve this, \sys can use BGP routing information or other network management tools (\eg traceroute data) to determine a subset of all deployed \mbx with maximally diverse paths to a particular source address. Then, the \mb initially receiving a request is responsible for redirecting the request to the set of pre-selected \mbx according to the source address in the request. Thus, if the sender can receive all tokens issued by each of these \mbx, it proves that the client is on-path for each of these \mbx, which, with higher probability, shows that the sender indeed owns its claimed source address. 

\parab{Source AS Assistance.} With the assistance of the source AS, the on-path attack can be bypassed entirely, ensuring that only a sender in the source AS has access to the token. Each supporting source AS would register an authentication proxy with \sys. To verify that the authentication proxy is approved by the source AS, the registration is signed using the rPKI certificates~\cite{rPKI} that prove control of both the IP address prefix being registered and the associated AS number. Then, when an \mb receives a request from an address, it first determines whether that address is represented by an authentication proxy. If so, it establishes a key with that authentication proxy, using the public key from the rPKI certificates. The \mb then encrypts and provides the \{client source address, token\} pair to the authentication proxy, and redirect the client to fetch the token from its authentication proxy. Because the token is encrypted to the authentication proxy, on-path adversaries gain no additional information, and thus a sender can only claim a source address only with the source AS' approval. 

\parab{Client Support via Multi-Context TLS.} This source authentication mechanism, based on mcTLS~\cite{mcTLS}, requires lightweight upgrade (\eg software installation) from clients. However, it offers at least two advantages. First, it enables the victim to explicitly control which portions of an encrypted connection can be read or write by an \mb (without sharing any credentials with the cloud provider hosting the \mb). For instance, the victim can explicitly allow the \mb to read the HTTP Host header (carrying the token) so that the \mb can perform source authentication directly without having to relay the encrypted connection to the victim any more, which addresses the second weakness of our strawman design. Second, with such lightweight client-end upgrade, it is easier for \sys to identify the premium clients preferred by the victim. In particular, after a premium client successfully authenticates itself (for instance through dedicated authentication services) to \sys that the client is indeed a premium client for the victim, \sys can issue the client a cryptographically signed pseudonym. For future requests, the premium client can carry its pseudonym in a customized HTTP header (\eg \textsf{X-Pseudonym}) to allow \mbx to properly identity its requests. 

%% file: implementation.tex
\section{Implementation}\label{sec:implementation}
\begin{figure}[t]
	\centering
	\mbox{
		\includegraphics[width=0.98\columnwidth]{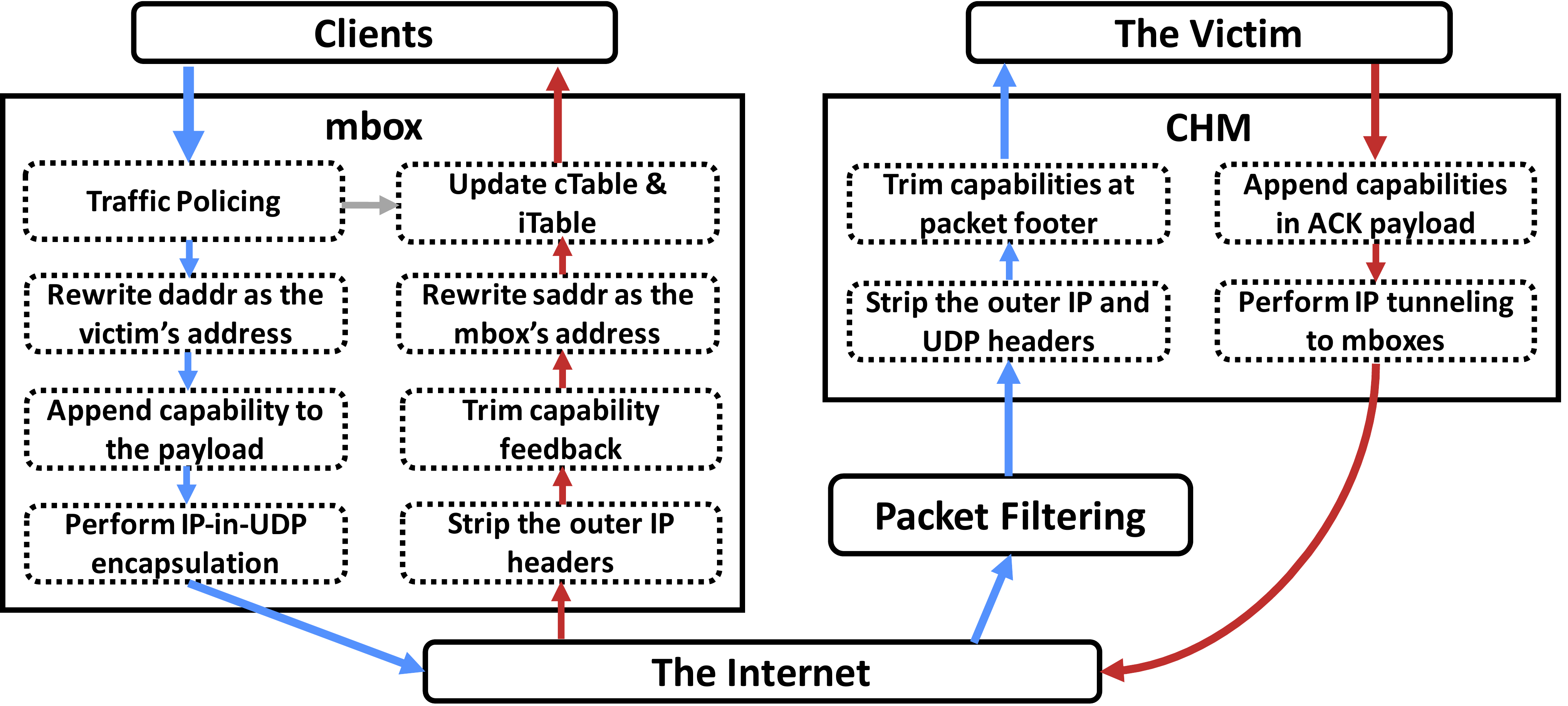}
	}
	\caption{The software stack of the \mb and CHM.}
	\label{fig:software_stack}
\end{figure}

We present the implementation of \sys's prototype in this section. The source code is publicly available at Github~\cite{source_code}.

\subsection{The Implementation of \mbx and CHM}
The \mbx and the CHM at the victim are implemented based on the NetFilter Linux Kernel Module, which combined have ${\sim}1500$ 
lines of $\texttt{C}$ code (excluding the capability generation code). The software stack of our implementation is illustrated in Figure~\ref{fig:software_stack}.  

All inbound traffic from clients to an \mb is subject to the traffic policing whereas only accepted packets go through the capability-related processing. Packet dropping due to traffic policing triggers \itable updates.  For each accepted packet, the \mb rewrites its destination address as the victim's address to point the packet to the victim. To carry capabilities, rather than defining a new packet header, the \mb appends the capabilities to the end of the original data payload, which avoids compatibility problems at intermediate routers and switches. The CHM is responsible for trimming these capabilities to deliver the original payload to the victim's applications. If the packet filter is deployed, the \mb performs the IP-in-UDP encapsulation, and uses the UDP source and destination port number to carry an authenticator. All checksums need to be recomputed after packet manipulation to ensure correctness. ECN and encapsulation interactions are addressed in~\cite{RFC6040}. 

To avoid packet fragmentation due to the additional $68$ bytes added by the \mb ($20$ bytes for outer IP header, $8$ bytes for the outer UDP header, and $40$ bytes reserved for a capability), the \mb needs to be \emph{a priori} aware of the MTU $M_d$ on its path to the victim. Then the \mb sets its MSS  to no more than $M_d{-}68{-}40$ (the MSS is 40 less than the MTU). We do not directly set the MTU of the \mb's NIC to rely on the path MTU discovery to find the right packet size because some ISPs may block ICMP packets. On our testbed, $M_d=1500$, so we set the \mb's MSS to $1360$. 

Upon receiving packets from upstream \mbx, the CHM strips their outer IP/UDP headers and trims the capabilities. To return these capabilities, the CHM piggybacks capabilities to the payload of ACK packets. To ensure that a capability is returned to the \mb issuing the capability even if the Internet path is asymmetric, the CHM performs IP-in-IP encapsulation to tunnel the ACK packets to the right \mb. We allow one ACK packet to carry multiple capabilities since the victim may generate cumulative ACKs rather than per-packet ACKs. Further, the CHM tries to pack more capabilities in one ACK packet to reduce the capability feedback latency at the CHM. The number of capabilities carried in one ACK packet is stored in the TCP option (the $4$-bit \texttt{res1} option). Thus, the CHM can append up to $15$ capabilities in one ACK packet if the packet has enough space and the CHM has buffered enough capabilities. 

Upon receiving an ACK packet from the CHM, the \mb strips the outer IP header and trims the capability feedback (if any) at the packet footer. Further, the \mb needs to rewrite the ACK packet's source address back to its own address, since the client's TCP connection is expecting to communicate with the \mb. Based on the received capability feedback, the \mb updates the \itable and \ctable accordingly to support the traffic policing algorithm.

\subsection{Capability Generation}
We use the AES-128 based CBC-MAC, based on the Intel AES-NI library, to compute MACs, due to its fast speed and availability in modern CPUs~\cite{aes,aes_intel}. We port the capability implementation (${\sim}400$ lines of \texttt{C} code) into the \mb and CHM 
kernel module. The \mb needs to perform both capability generation and verification whereas the CHM performs only verification. 

%% file: evaluation.tex
\section{Evaluation}\label{sec:evaluation}
\subsection{The Internet Experiments}\label{sec:evaluation:Internet}
This section studies the path length and latency inflation for rerouting clients' traffic to \mbx hosted in the cloud. 

\subsubsection{Path Inflation}  
We construct the AS level Internet topology based on the CAIDA AS relationships dataset~\cite{AS-dataset}, including 52680 ASes and their business relationships~\cite{AS-relationship}. To construct the communication route, two constraints are applied based on the routes export policies in~\cite{AS-economic1,AS-economic2}. First, an AS prefers customer links over peer links and peer links over provider links. Second, a path is valid only if each AS providing transit is paid. Among all valid paths, an AS prefers the path with least AS hops (a random tie breaker is applied if necessary). As an example, we use Amazon EC2 as the cloud provider to host the \mbx, and obtain its AS number based on the report~\cite{AS_name}. Amazon claims $11$ ASes in the report. We first exclude the ASes not appearing in the global routing table, and find that AS 16509 is the provider for the remaining Amazon ASes, so we use AS 16509 to represent Amazon.  

\def\hop{\text{\it hop}}
\def\infla{\text{\it infla}}
\def\short{\text{\it short}}
\def\cut{\text{\it cut}}
\def\no{\text{\it no}}

We randomly pick 2000 ASes as victims, and for each victim we randomly pick 1500 access ASes. Among all victims, 1000 victims are stub ASes without direct customers and the remaining victims are non-stub ASes. For each AS-victim pair, we obtain the direct route from the access AS to the victim, and the rerouted path through an \mb. Table~\ref{tab:hop_status} summarizes the route comparison. $N^{\hop}_{\infla}$ is the average AS-hop inflation of the rerouted path compared with the direct route. $P^{\short}_{\cut}$ is the percentage of access ASes that can reach the victim with fewer hops after rerouting and $P^{\no}_{\infla}$ is percentage of ASes without hop inflation.

\begin{table}[t] 
	\centering
	\begin{tabular}{|>{}c|>{}c|>{}c|>{}c|>{}c|>{}c|>{}c|>{}c|>{}c|>{}c|}
		\hline
		Victims&
		$N^{\hop}_{\infla}$&
		$P^{\short}_{\cut}$ &
		$P^{\no}_{\infla}$\\
		\hline
		Non-stub ASes&
		1.1&
		10.6\%&
		22.2\%\\
		\hline
		Stub ASes&
		1.5&
		8.4\%&
		18.0\%\\
		\hline
		Overall&
		1.3&
		9.5\%&
		20.1\%\\
		\hline
	\end{tabular}
	\caption{Rerouting traffic to \mbx causes small AS-hop inflation, and ${\sim}10$\% access ASes can 
	even reach the victim with fewer hops through \mbx.}
	\label{tab:hop_status}
\end{table}

\begin{figure}[t]
	\centering
	\mbox{
		\includegraphics[width=0.8\columnwidth]{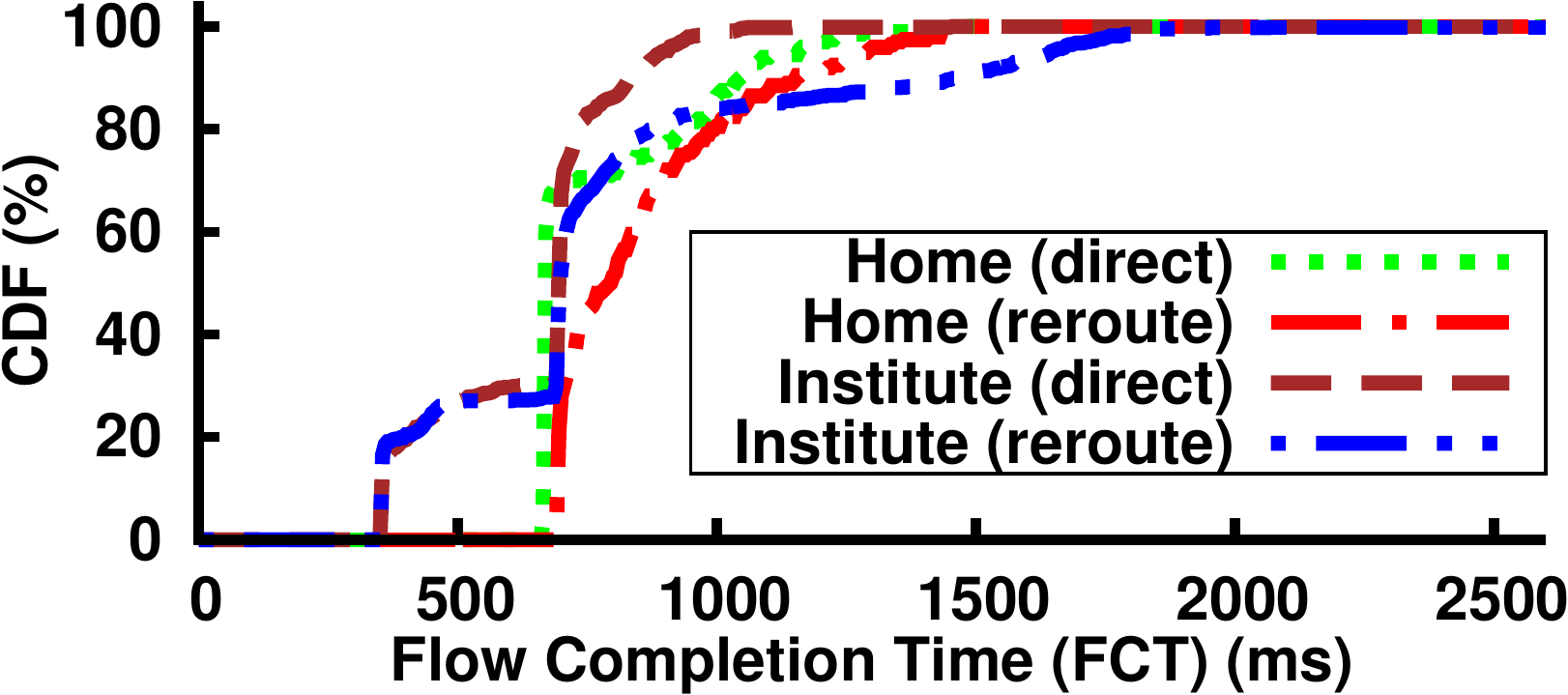}
	}
	\caption{[Internet] FCTs for 
		direct paths and rerouted paths  
		under various Internet conditions.}
	\label{fig:fct}
\end{figure}

Overall, it takes an access AS $1.3$ more AS-hops to reach the victim after rerouting. Even for stub victims, which are closer the Internet edge, the average hop inflation is only $1.5$. We also notice that ${\sim}10$\% ASes have \emph{shorter} paths due to the rerouting.  

Besides EC2, we also perform path inflation analysis when \mbx are hosted by CloudFlare. The results show that the average path inflation is about $2.3$ AS-hops. For any cloud provider, \sys has the \emph{same path inflation} as the cloud-based DDoS solutions hosted by the same cloud provider, since capability feedback is carried in ACK packets. As such, deploying \sys into existing cloud-based systems does not increase path inflation.

\subsubsection{Latency Inflation} 
In this section, we study the latency inflation caused by the rerouting. In our prototype running on the Internet, we deploy $3$ \mbx on Amazon EC2 (located in North America, Asia and Europe), one victim server in a US university and about one hundred senders (located in North America, Asia and Europe) on PlanetLab~\cite{planetlab} nodes. We also deploy few clients on personal computers to test \sys in home network. The wide distribution of clients allows us to evaluate \sys on various Internet links. We did not launch DDoS attacks over the Internet, which raises ethical and legal concerns. Instead, we evaluate how \sys may affect the clients in the normal Internet without attacks, and perform the experiments involving large scale DDoS attacks on our private testbed and in simulation. 

In the experiment, each client posts a $100$KB file to the server, and its traffic is rerouted to the nearest \mb before reaching the server. We repeat the posting on each client 10,000 times to reduce sampling error. We also run the experiment during both peak hours and midnight (based on the server's timezone) to test various network conditions. As a control, clients post the files to server via direct paths without traversing though \mbx. 

Figure \ref{fig:fct} shows the CDF of the flow completion times (FCTs) for the file posting. Overall, we notice ${\sim}9$\% average FCT inflation, and less than $5$\% latency inflation in home network. Therefore, traffic rerouting introduces small extra latency to the clients. 

\sys's latency inflation includes both rerouting-induced networking latency and capability-induced computational overhead. As evaluated in \S~\ref{sec:evaluation:overhead}, the per-packet processing latency overhead caused by capability computation is ${\sim}1.4~\mu$s, which is negligible compared with typical Internet RTTs. Thus, \sys has latency almost identical to the existing cloud-based DDoS mitigation service.

\subsection{Testbed Experiments}\label{sec:evaluation:testbed}

\begin{figure}[t]
	\centering
	\mbox{
		\includegraphics[width=0.95\columnwidth]{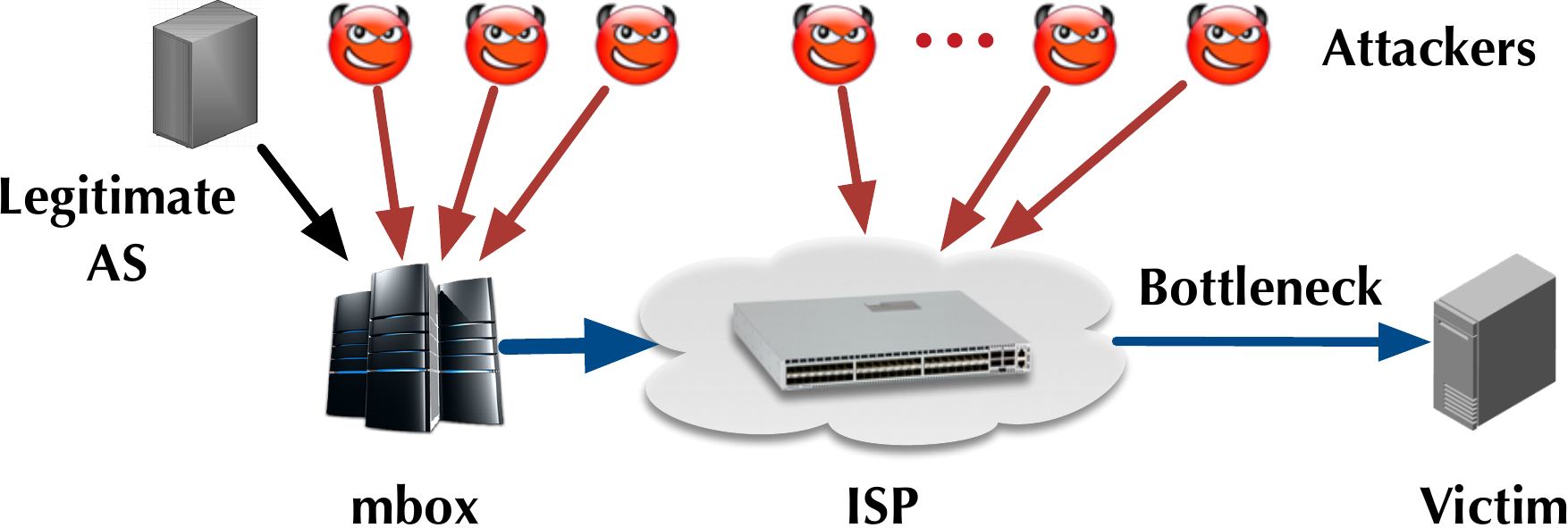}
	}
	\caption{Testbed network topology.}
	\label{fig:testbed}
\end{figure}

\begin{figure}[t]
	\centering
	\mbox{
		\includegraphics[width=0.95\columnwidth]{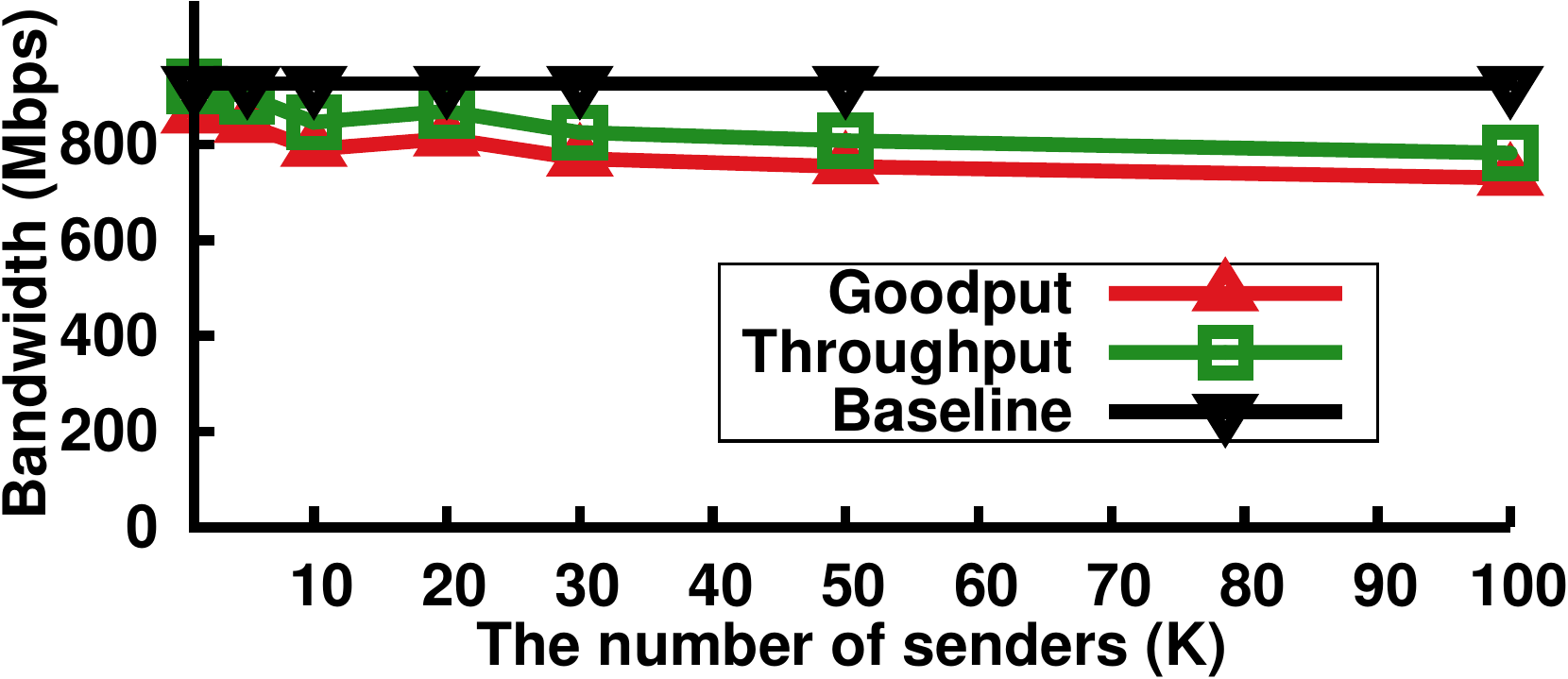}
	}
	\caption{[Testbed] Throughput and goodput when \sys policies   
		different numbers of senders.}
	\label{fig:overhead}
\end{figure}

\subsubsection{Traffic Policing Overhead} \label{sec:evaluation:overhead}
In this section, we evaluate the traffic policing overhead on our testbed. We organize three servers as one sender, one \mb and one receiver. All servers, shipped with a quad-core Intel $2.8$GHz CPU, run the $3.13.0$ Linux kernel. The \mb is installed with multiple Gigabit  NICs to connect both the sender and receiver. A long TCP flow is established between the sender and receiver, via the \mb,   to measure the throughput achieved by the sender and receiver. To emulate a large number of sources, the \mb creates an \itable with $N$ entries. Each packet from the sender triggers a table look up for a random entry. We implement a two-level hash table in the kernel space to reduce the look up latency. Then the \mb generates a capability based on the obtained entry.

Figure \ref{fig:overhead} shows the measured throughput and goodput under various $N$. The goodput is computed by subtracting the additional header and capability size from the total packet size. The baseline throughput is obtained without \sys. Overall, the policing overhead in high speed network is small. When a single \mb deals with 100,000 sources sending \emph{simultaneously}, throughput drops by ${\sim}10$\%. Equivalently, \sys adds around $1.4$ microseconds latency to each packet processed. By replicating more geographically distributed \mbx in the cloud, the victim can distribute the workload across many \mbx when facing large scale attacks. 

\subsubsection{Enforce Destination-Defined Traffic Control Policies}
We now evaluate \sys's performance for enforcing victim-defined traffic control policies, along with the effectiveness of filtering bypassing traffic. This section evaluates \emph{pure} \sys. In reality, once incorporated by existing cloud-based DDoS prevention vendors, \sys needs only to process traffic that passes their pre-deployed defense. 

\parab{Testbed Topology.} Figure~\ref{fig:testbed} illustrates the network topology, including a single-homed victim AS purchasing $1$Gbps bandwidth from its ISP, an \mb and $10$ access ASes. The ISP is emulated by a Pronto-3297 48-port Gigabit switch to support packet filtering. The \mb is deployed on a server with multiple Gigabit NICs, and each access AS is deployed on a server with a single NIC. We add $100$ms latency at the victim via Linux traffic control to emulate the typical Internet RTT. To emulate large scale attacks, $9$ ASes are compromised. Attackers adopt a hybrid attack profile: $6$ attack ASes directly send large volumes of traffic to the victim, emulating amplification-based attacks, and the remaining attack ASes route traffic through the \mb. Thus, the total volume of attack traffic is $9$ times as much as the victim's bottleneck link capacity. Both the inbound and outbound points of the \mb are provisioned with $4$Gbps bandwidth to ensure   the \mb is not the bottleneck, emulating that the \mb is hosted by the well-provisioned cloud.  

\begin{figure}[t]
	\centering
	\mbox{
		\includegraphics[width=0.98\columnwidth]{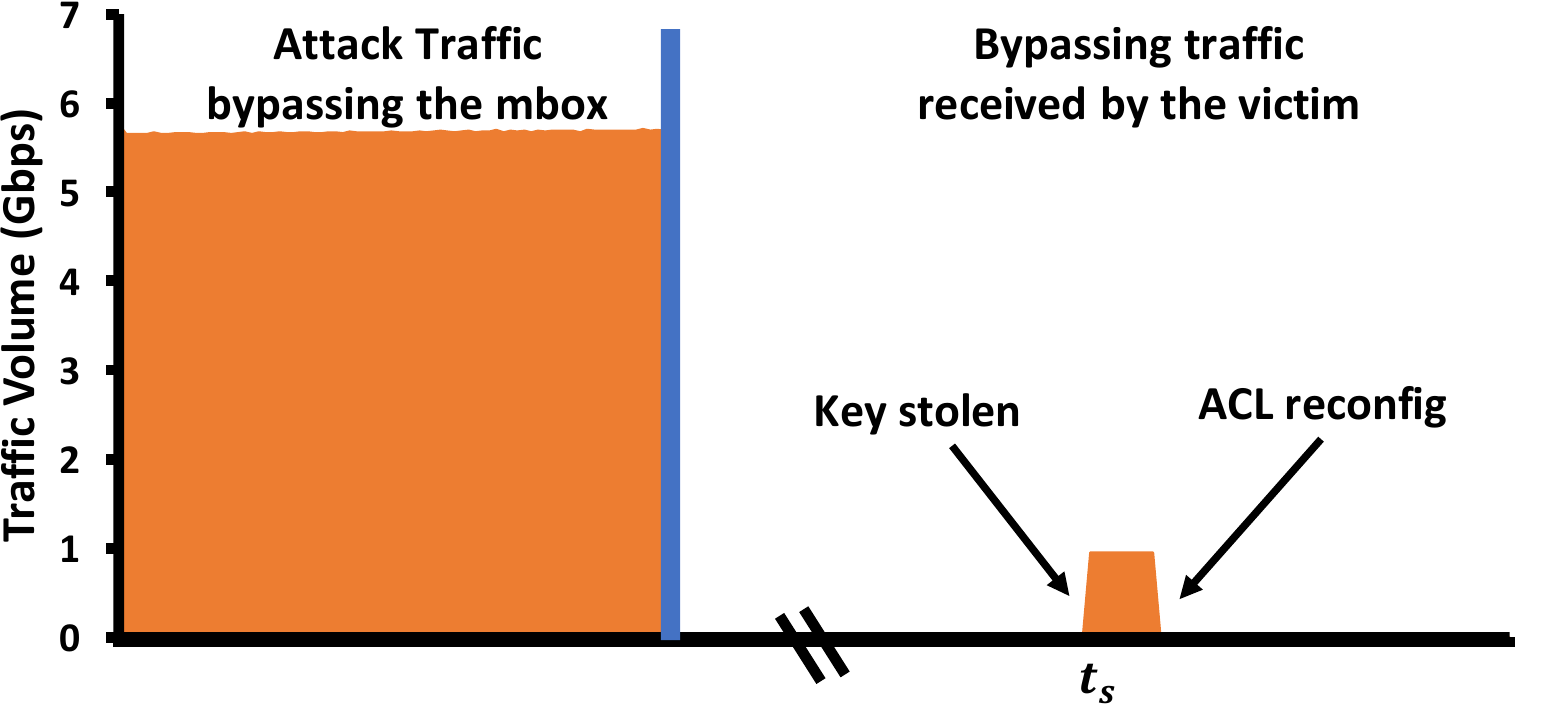}
	}
	\caption{[Testbed] Packet filtering via ACL.}
	\label{fig:filtering}
\end{figure}

\parab{Packet Filtering.} We first show the effectiveness of the packet filter. Six attack ASes spoof the \mb's source address and 
send $6$Gbps UDP traffic to the victim. The attack ASes scan all possible UDP port numbers to guess the shared secret. Figure~\ref{fig:filtering} shows the volume of attack traffic bypassing the \mb and its volume received by the victim. As the chance of a correct guess is very small, the filter can effectively stop the bypassing traffic from reaching the victim. Further, even if the shared secret were stolen by attackers at time $t_s$, the CHM would suddenly receive large numbers of packets without valid capabilities. Since packets traversing the \mb carry capabilities, the CHM realizes that the upstream filtering has been compromised. The victim then re-configures the ACL using a new secret to recover from key compromise. The ACL is effective within few milliseconds after reconfiguration. Thus, the packet filtering mechanism can promptly react to a compromised secret.  

\addtolength{\floatsep}{-5pt}
\addtolength{\abovecaptionskip}{-3pt}

\begin{figure*}[t]
	\centering
	\mbox{
		\subfigure[\small Window sizes in flat-attacks.\label{fig:natural:testbed:a}]{\includegraphics[scale=0.25]{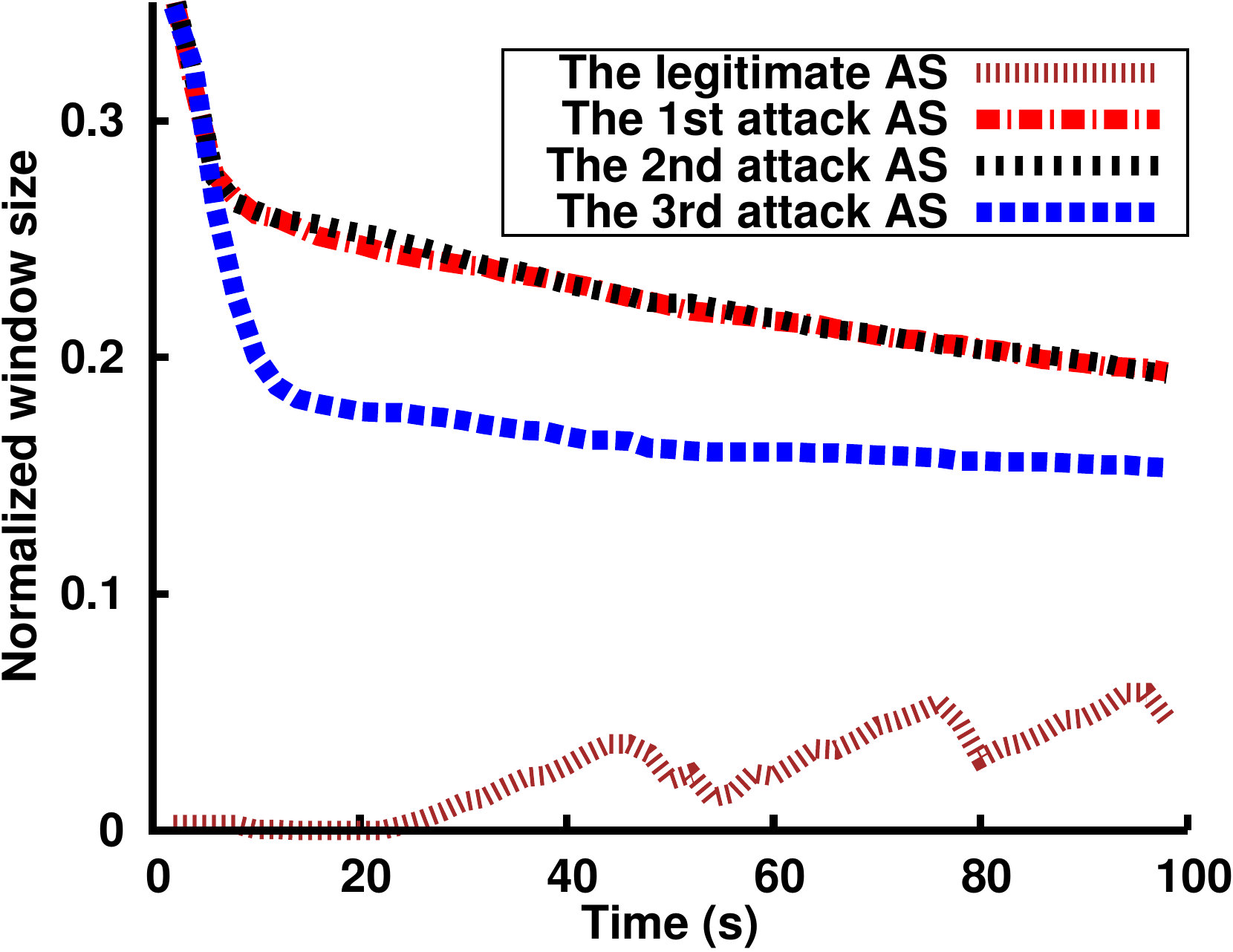}}
		\subfigure[\small LLRs in flat-attacks.\label{fig:natural:testbed:b}]{\includegraphics[scale=0.25]{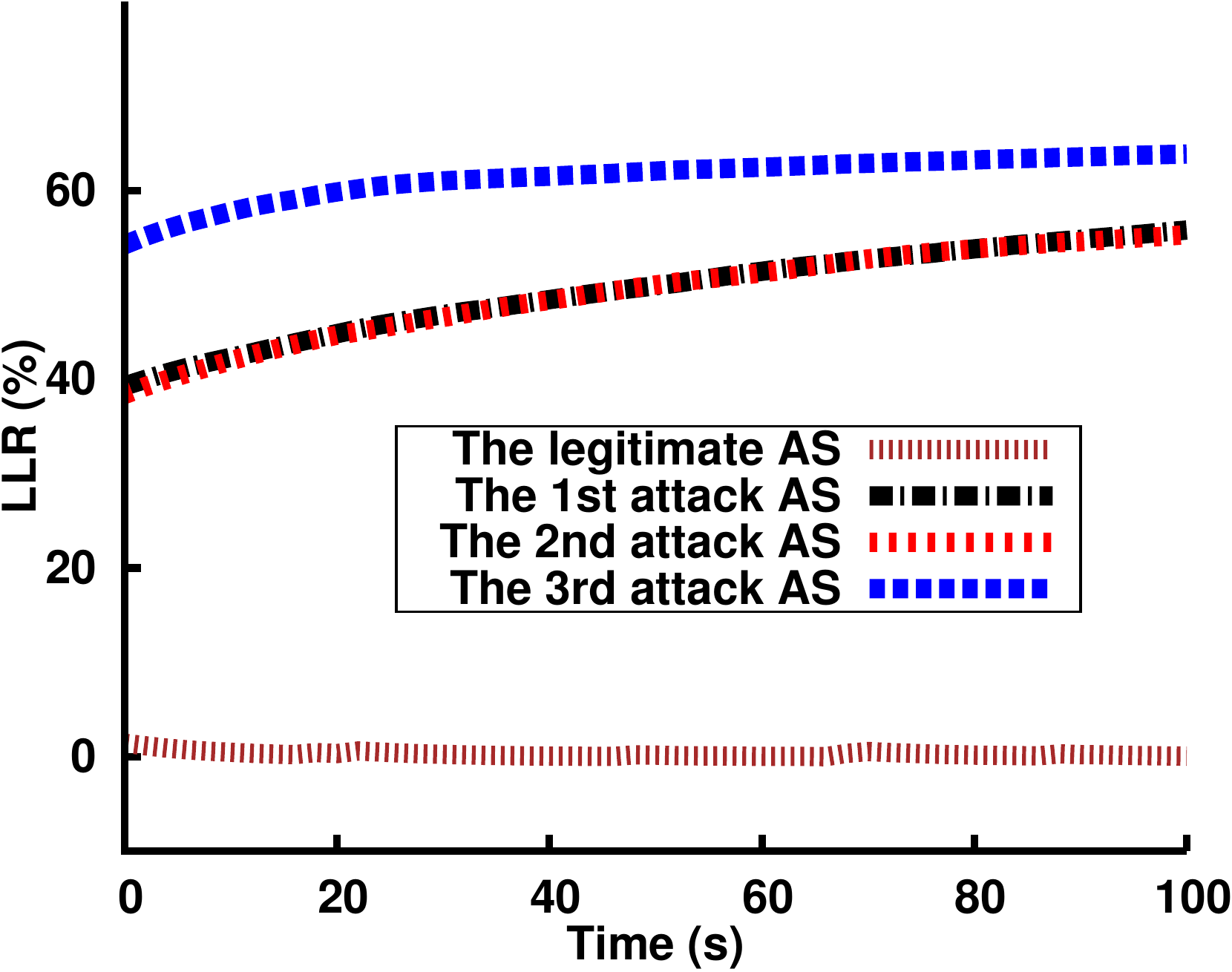}}
		\subfigure[\small Window sizes in reactive att.\label{fig:natural:testbed:c}]{\includegraphics[scale=0.25]{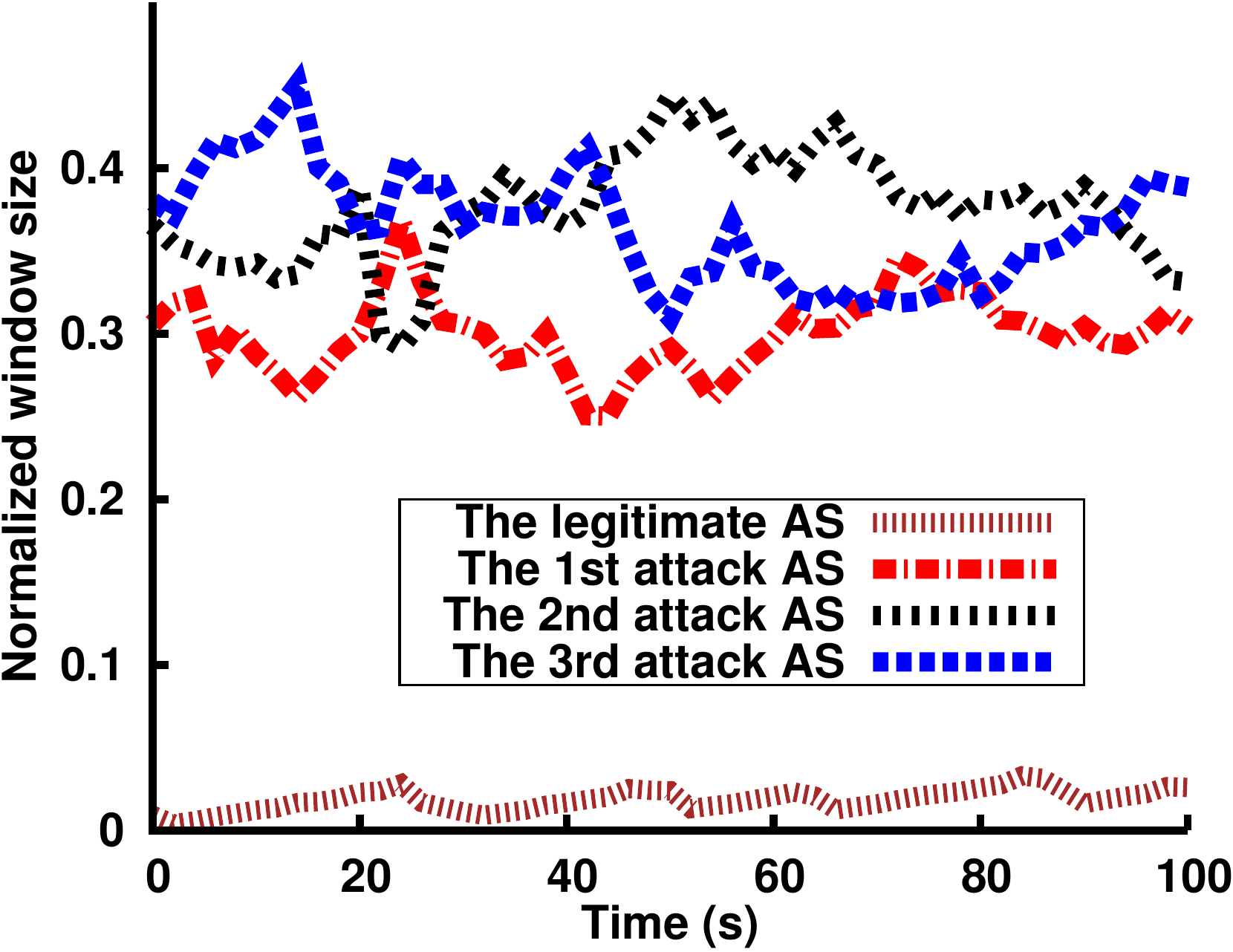}}
		\subfigure[\small LLRs in reactive attacks. \label{fig:natural:testbed:d}]{\includegraphics[scale=0.25]{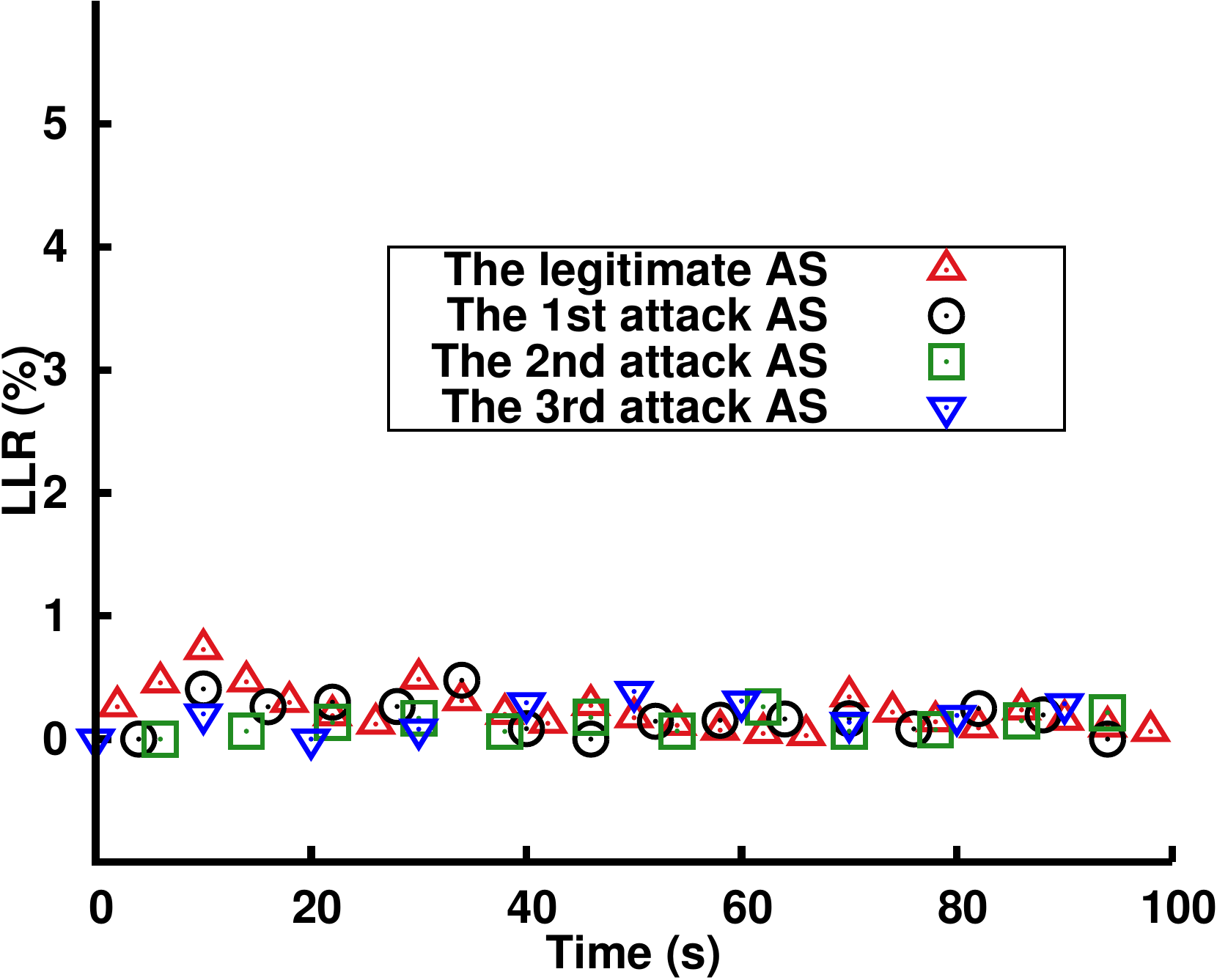}}
	}
	\caption{[Testbed] Enforcing the \textsf{NaturalShare}  
		policy. The legitimate AS gradually obtains a certain 
		amount of bandwidth under flat-rate attacks since attackers' 
		window sizes drop consistently over time 
		(Figure \ref{fig:natural:testbed:a}) due to their  
		high LLRs (Figure \ref{fig:natural:testbed:b}). However, 
		the attack ASes can consume over 95\% of the 
		bottleneck bandwidth via reactive attacks (Figure \ref{fig:natural:testbed:c}) while 
		maintaining low LLRs similar to the legitimate AS's LLR
		(Figure \ref{fig:natural:testbed:d}).}
	\label{fig:natural:testbed}
\end{figure*}

\begin{figure*}[t]
	\centering
	\mbox{
		\subfigure[\small Window sizes in flat-attacks.\label{fig:perAS:testbed:a}]{\includegraphics[scale=0.25]{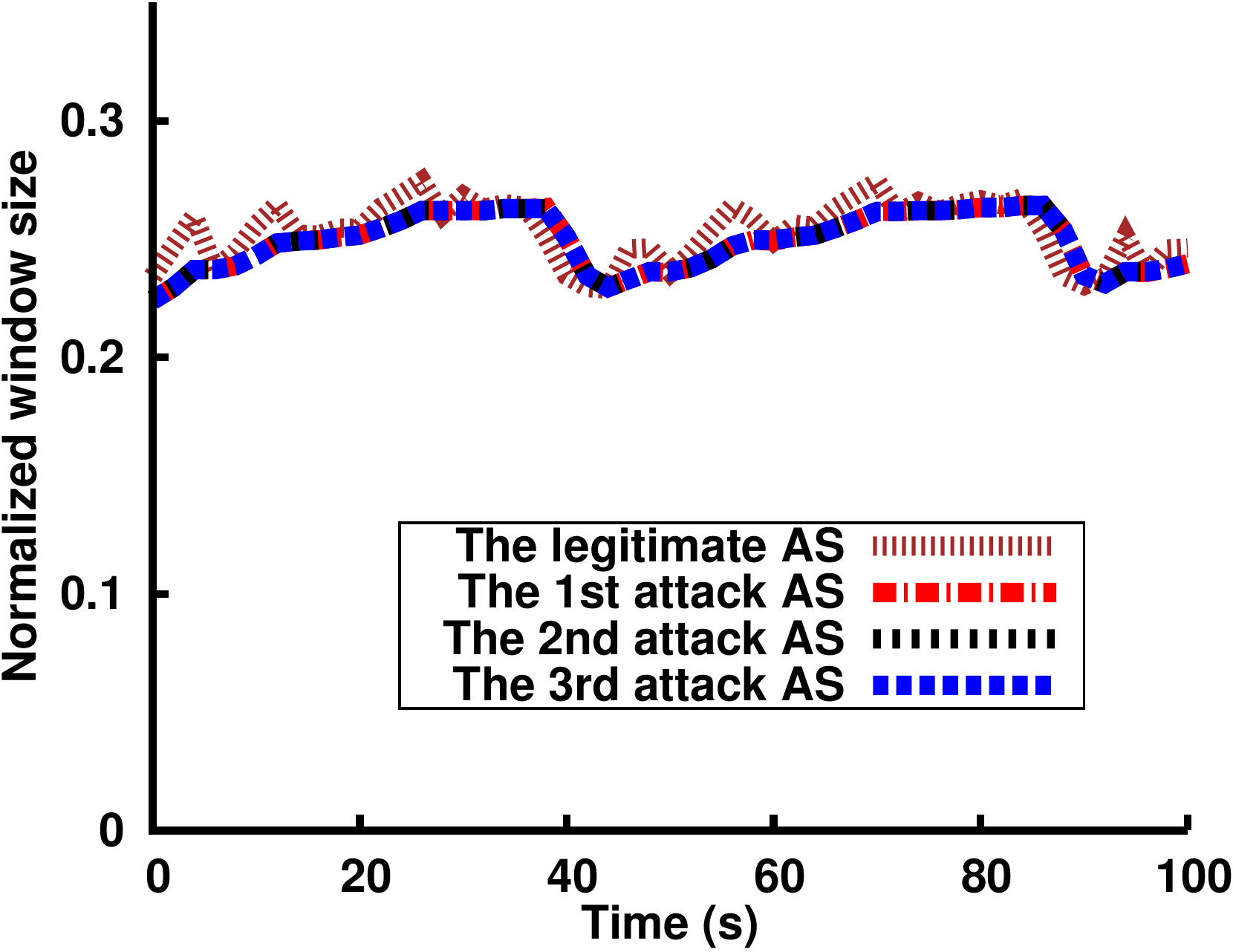}}
		\subfigure[\small LLRs in flat-attacks.\label{fig:perAS:testbed:b}]{\includegraphics[scale=0.25]{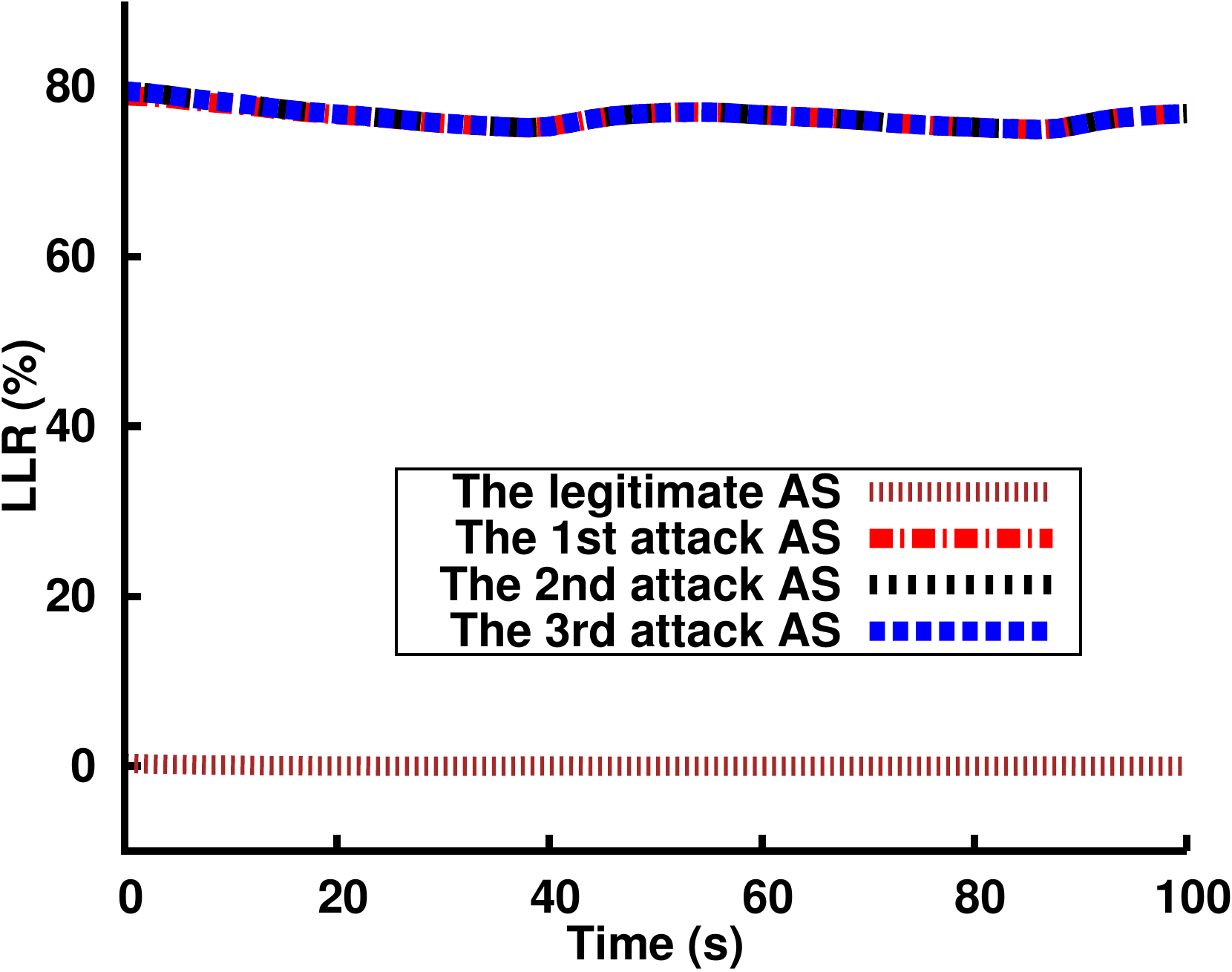}}
		\subfigure[\small Window sizes in reactive att.\label{fig:perAS:testbed:c}]{\includegraphics[scale=0.25]{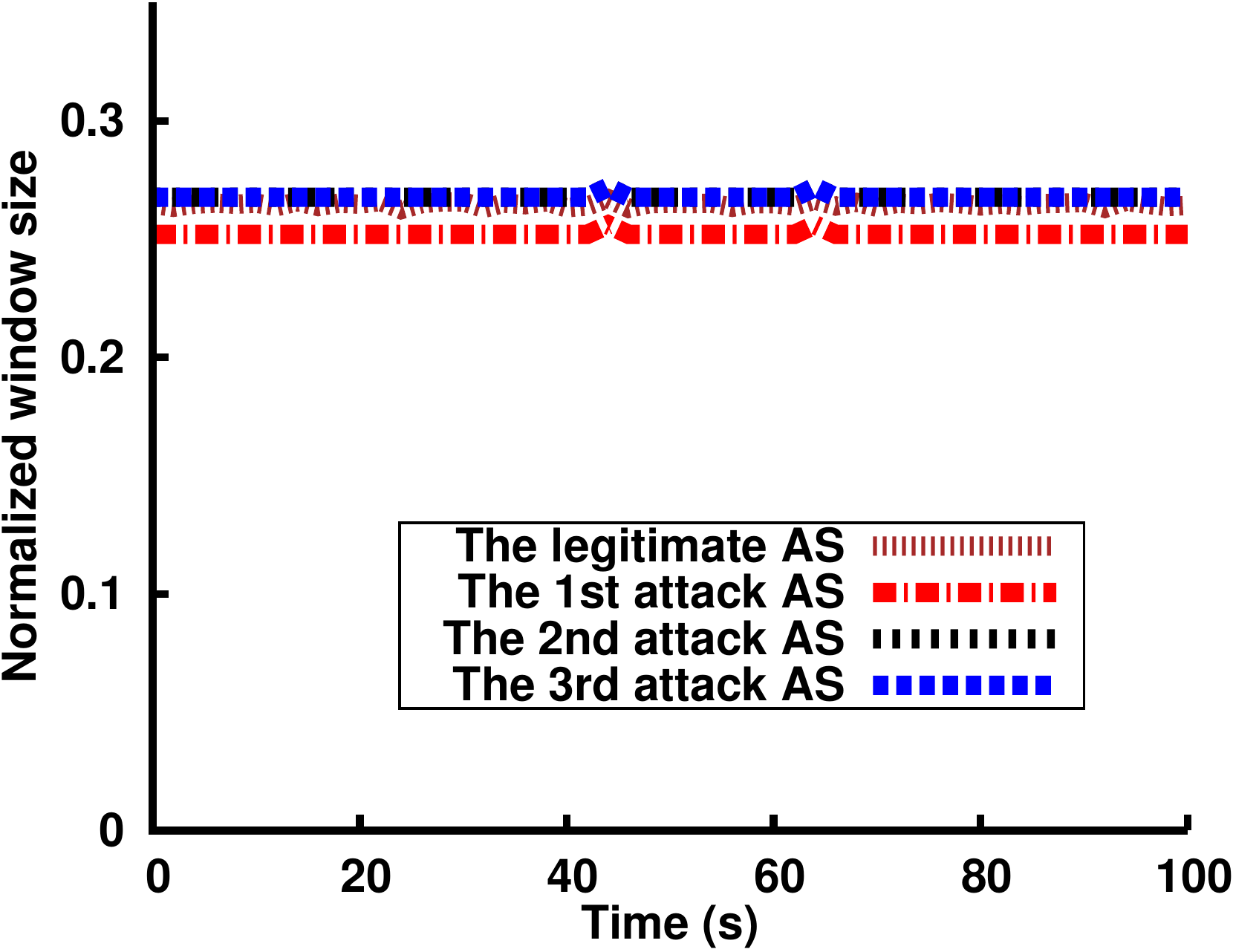}}
		\subfigure[\small LLRs in reactive attacks. \label{fig:perAS:testbed:d}]{\includegraphics[scale=0.25]{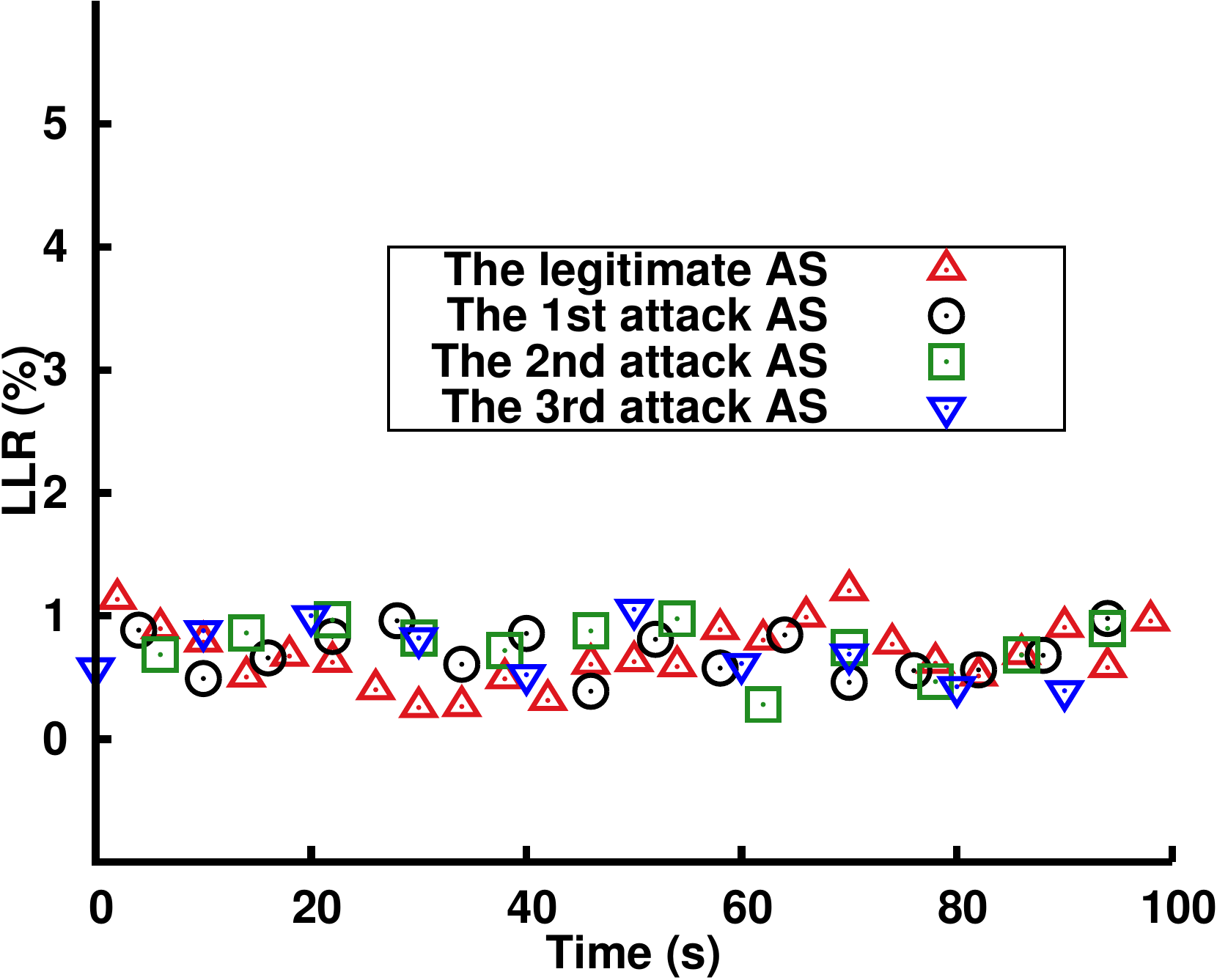}}
	}
	\caption{[Testbed] Enforcing the \textsf{PerASFairshare}  
		policy. The legitimate AS can obtain 
		at least the per-AS fair rate at the bottleneck 
		regardless of the attack strategies (Figures 
		\ref{fig:perAS:testbed:a} and \ref{fig:perAS:testbed:c}). Further, 
		the legitimate AS gains slightly more bandwidth than 
		the attackers under flat-rate attacks as the attack ASes have large LLRs (Figure 
		\ref{fig:perAS:testbed:b}).}
	\label{fig:perAS:testbed}
\end{figure*}

\parab{\textsf{NaturalShare} and \textsf{PerASFairshare} Policies.} In this section, we first show that \sys can enforce both  \textsf{NaturalShare} and \textsf{PerASFairshare} policies described in \S~\ref{sec:bandwidt_sharing_policy}. We use the default parameter setting in Table~\ref{tab:para}, and defer detailed parameter study in \S~\ref{sec:evaluation:large_scale}. Since \sys conditionally allows an AS to send faster than its $\mathcal{W}_R$, we use the \emph{window size}, defined as the larger value between an AS's $\mathcal{W}_R$ and its delivered packets to the victim, as the performance metric. For clear presentation, we normalize the window size to the maximum number of $1.5$KB packets deliverable through a $1$Gbps link in one detection period. We do not translate window sizes to throughput because packet sizes vary. 

Attackers adopt two representative strategies: \first they send flat rates regardless of packet losses, and \second they dynamically adjust their rates based on packet losses (reactive attacks). To launch flat-rate attacks, the attackers keep sending UDP traffic to the victim. The CHM uses a dedicated flow to return received capabilities to the \mb since no ACK packets are generated for UDP traffic. One way of launching reactive attacks is that the attackers simultaneously maintain many more TCP flows than the legitimate AS. Such a many-to-one communication pattern allows the attackers to occupy almost the entire bottleneck, even through each single flow seems completely ``legitimate''.  

The legitimate AS always communicates with the victim via a long-lived TCP connection. 

Figure~\ref{fig:natural:testbed} shows the results for the \textsf{NaturalShare} policy. As the bottleneck is flooded by attack traffic, the legitimate AS is forced to enter  timeout at the beginning, as illustrated in Figure~\ref{fig:natural:testbed:a}. The attackers' window sizes are decreasing over time, which can be explained via Figure~\ref{fig:natural:testbed:b}. As the volume of attack traffic is well above the bottleneck's capacity, all attack ASes' LLRs are well above $\Th_{\slr}^{\drop}$. Thus, the \mb drops all their best-effort packets. As a result, when one attack AS's window size is $W(t)$ in detection period $t$, then $W(t+1) \leq W(t)$ since  in period $t{+}1$ any packet sent beyond $W(t)$ is dropped. Further, any new packet losses from the attack AS, caused by an overflow at the bottleneck buffer, will further reduce $W(t+1)$. Therefore, all attack ASes' window sizes are consistently decreasing over time, creating spare bandwidth at the bottleneck for the legitimate AS. As showed in Figure~\ref{fig:natural:testbed:a}, the legitimate AS gradually recovers from timeouts. 

\begin{figure}[t]
	\centering
	\mbox{
		\includegraphics[width=0.9\columnwidth]{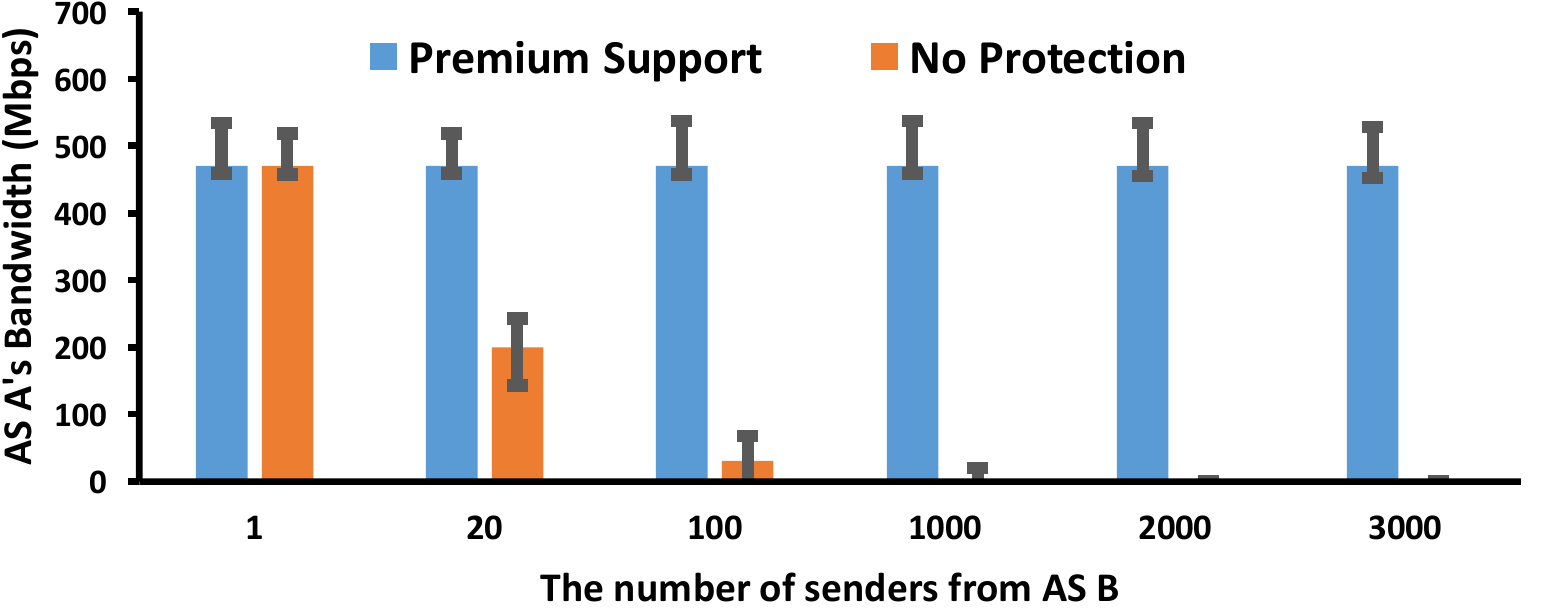}
	}
	\caption{[Testbed] \sys ensures that the premium client (AS A) receives consistent bandwidth.}
	\label{fig:premium}
\end{figure}

The \textsf{NaturalShare} policy, however, cannot well protect the legitimate AS if the attackers adopt the reactive attack strategy. By adjusting the sending rates based on packet losses, the attack ASes can keep their LLRs low enough to regain the advantage of delivering best-effort packets. Meanwhile, they can gain much more bandwidth by initiating more TCP flows. Figure~\ref{fig:natural:testbed:c} shows the window sizes when each attack AS starts $200$ TCP flows whereas the legitimate AS has only one. The attackers consume over $95\%$ of the bottleneck bandwidth, while keeping low LLRs similar to that of the legitimate AS (Figure~\ref{fig:natural:testbed:d}).

Figure~\ref{fig:perAS:testbed} shows the results for the \textsf{PerASFairshare} policy. Figures~\ref{fig:perAS:testbed:a} and \ref{fig:perAS:testbed:c} demonstrate that the legitimate AS receives  at least per-AS fair rate at the bottleneck regardless of the attack strategies, overcoming the shortcomings of the \textsf{NaturalShare} policy. Further, under flat-rate attacks, the legitimate AS has slightly larger window sizes than the attackers since, again, the \mb does not accept any best-effort packets from the attackers due to their high LLRs (as showed in Figure~\ref{fig:perAS:testbed:b}).

\begin{figure*}[t]
	\centering
	\mbox{
		\subfigure[\small \textsf{NaturalShare}. \label{fig:fairness_regime:a}]{\includegraphics[scale=0.26]{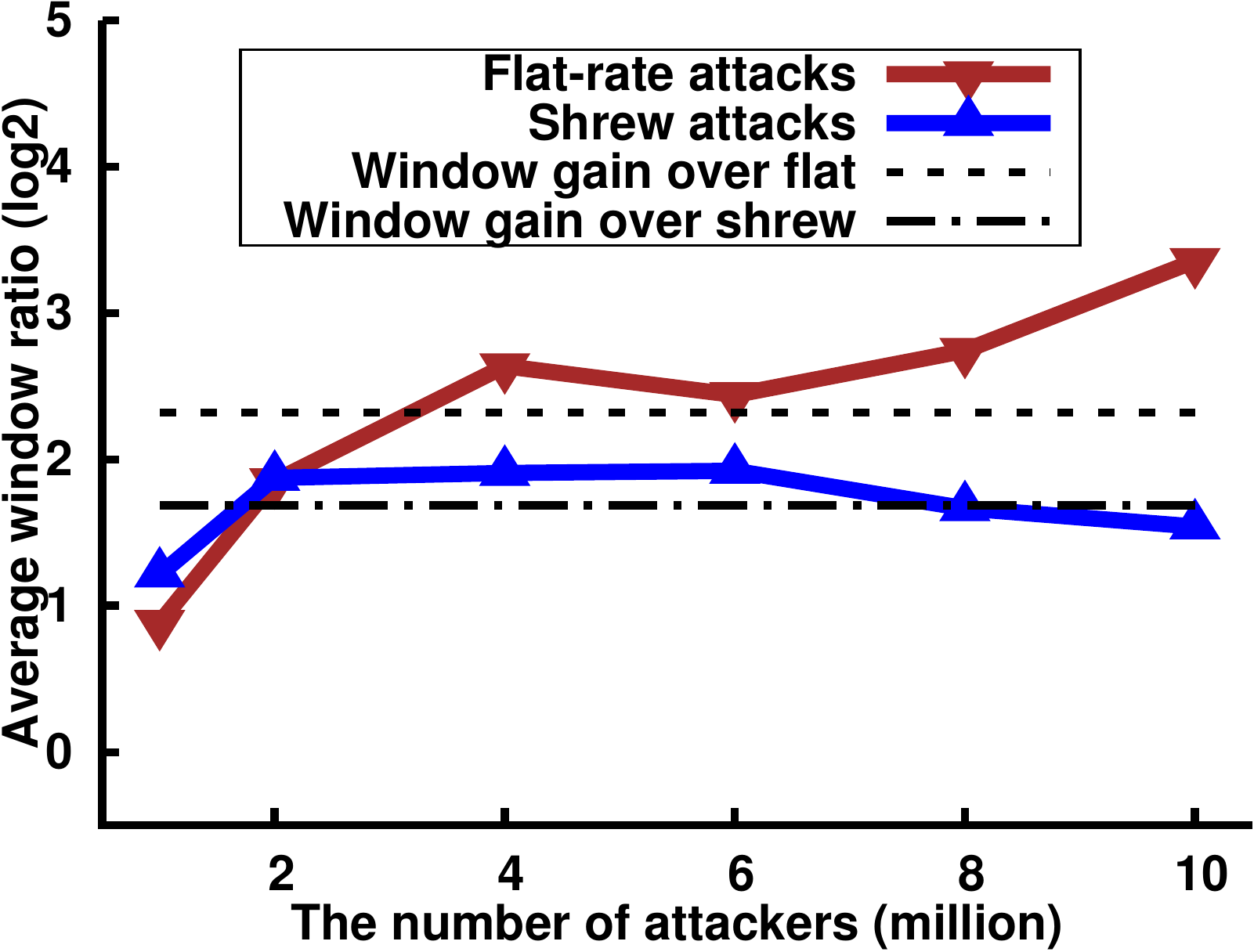}}
		\subfigure[\small \textsf{PerSenderFairshare}. \label{fig:fairness_regime:b}]{\includegraphics[scale=0.26]{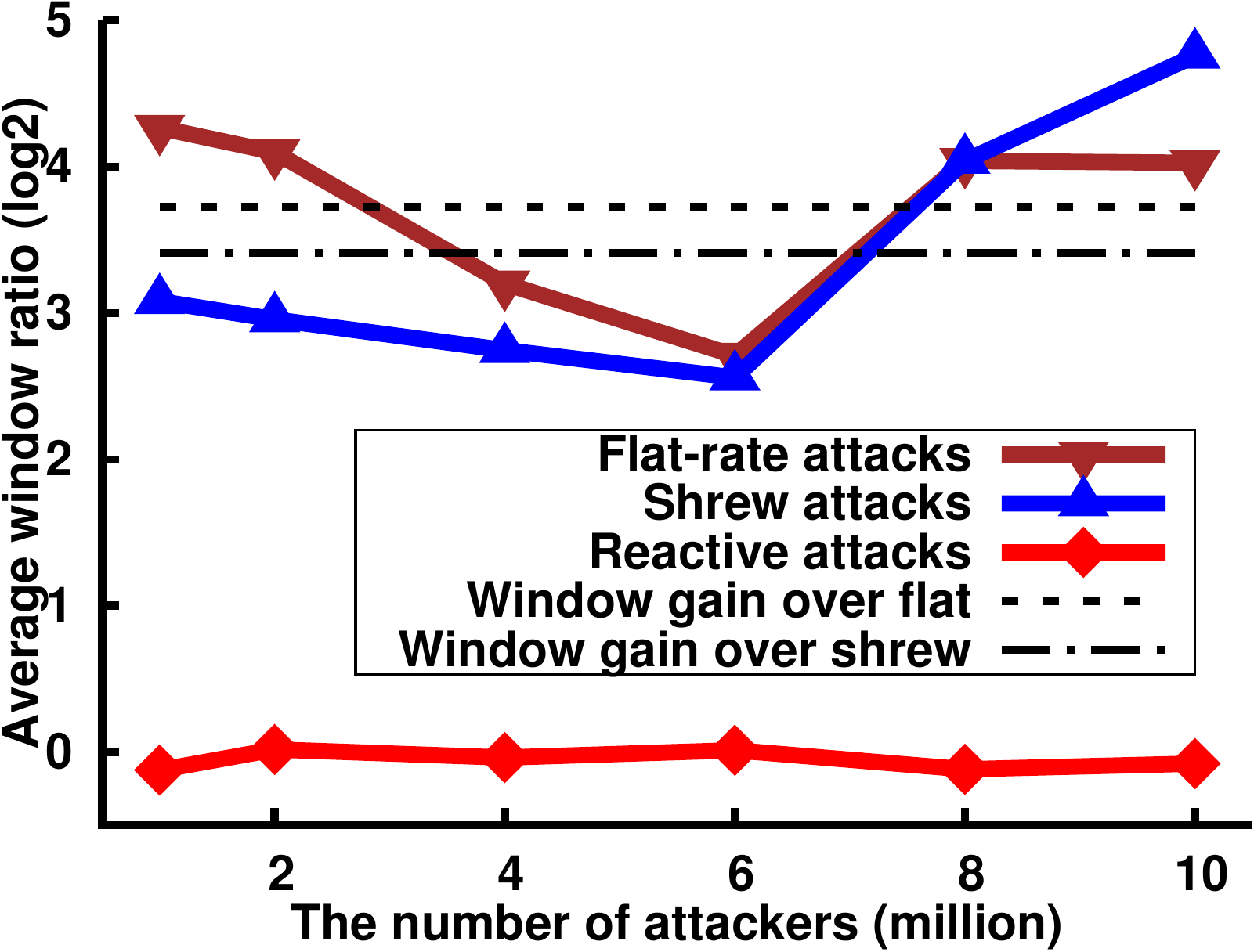}}
		\subfigure[\small Jain's fairness index (FI). \label{fig:fairness_regime:c}]{\includegraphics[scale=0.26]{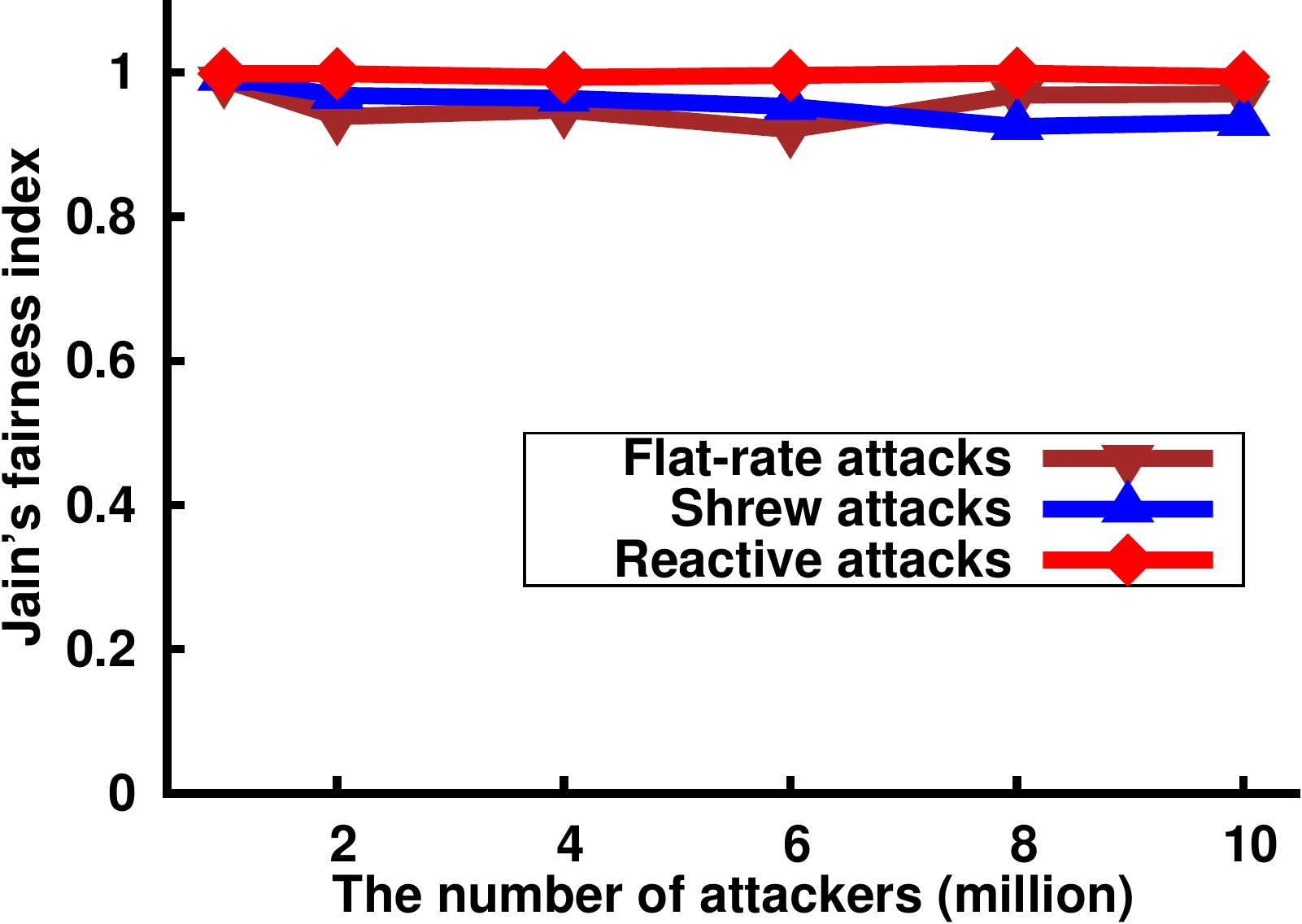}}
		\subfigure[\small FI for various \mb counts. \label{fig:fairness_regime:d}]{\includegraphics[scale=0.26]{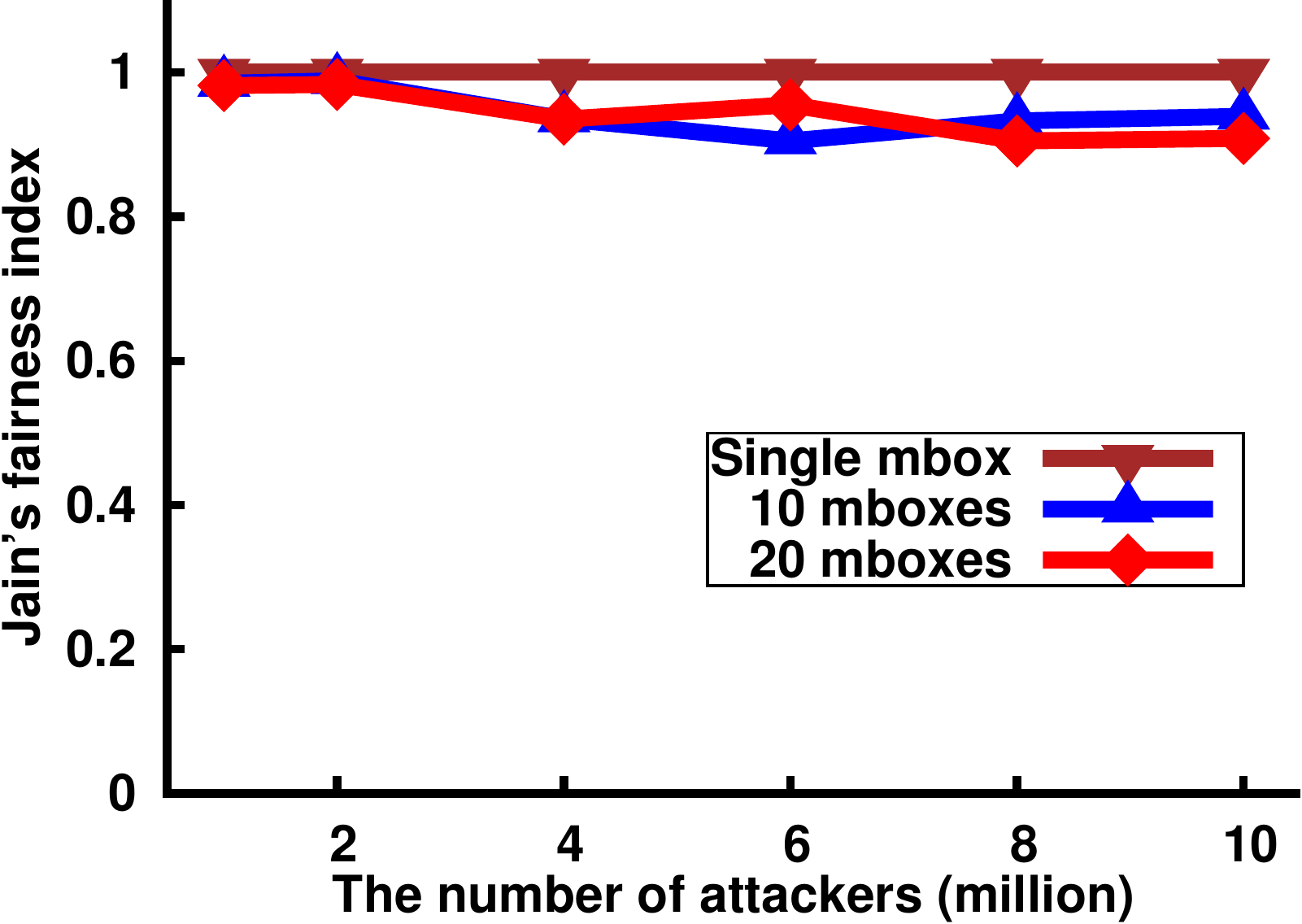}}
	}
	\caption{[Simulation] Evaluating \textsf{NaturalShare} \& \textsf{PerSenderFairshare} in large scale. 
		Figures \ref{fig:fairness_regime:a} and \ref{fig:fairness_regime:b} show that   
		the clients' average window size is larger than that of the attackers 
		under both flat-rate and shrew attacks. Figure \ref{fig:fairness_regime:c} 
		proves that the clients' window 
		sizes converge to fairness in the \textsf{PerSenderFairshare} policy.  
		Figure \ref{fig:fairness_regime:d} shows that 
		\sys can enforce strong fairness among all senders even without 
		coordination among the \mbx.}
	\label{fig:fairness_regime}
\end{figure*}

\parab{\textsf{PremiumClientSupport} Policy.} This section evaluates the \textsf{PremiumClientSupport} policy discussed in \S~\ref{sec:bandwidt_sharing_policy}. We consider a legitimate AS (AS A) that is a premium client which reserves half of the bottleneck bandwidth. Figure \ref{fig:premium} plots AS A's bandwidth when the number of senders from the attack ASes increases. With the \textsf{PremiumClientSupport} policy, \sys ensures AS A receives consistent bandwidth regardless of the number of senders from the attack ASes. However, without such a policy, the attack ASes can selfishly take away the majority of bottleneck bandwidth by involving more senders.  In reality, since an adversary can control large number of bots, the PremiumClientSupport policy turns out to be invaluable according to our interviews with industry people. 

\subsection{Large Scale Evaluation}\label{sec:evaluation:large_scale}
In this section, we further evaluate \sys via large scale simulations on ns-3 \cite{ns3}. We desire to emulate real-world DDoS attacks in which up to millions of bots flood a victim. To circumvent the scalability problem of ns-3 at such a scale, we adopt the same approach in NetFence \cite{netfence}, \ie by fixing the number of nodes (${\sim}5000$) and scaling down the link capacity proportionally, we can simulate attack scenarios where 1 million to 10 million attackers flood a 40Gbps link. The simulation topology is similar to the testbed topology, except that all attackers are connected to the \mb.

Besides the flat-rate attacks and reactive attacks, we also consider the on-off shrew attacks \cite{low-rate} in the simulations. Both the on-period and off-period in shrew attacks are 1s. The number of attackers is $10$ times larger than that of legitimate clients. In flat-rate attacks and shrew attacks, the attack  traffic volume is $3$ times larger than the capacity of the bottleneck. In reactive attacks, each attacker opens $10$ connections, whereas a client has one. The bottleneck router buffer size is determined based on~\cite{buffer_size}, and the RTT is $100$ms.  

\parab{\textsf{NaturalShare} \& \textsf{PerSenderFairshare} in Scale.} Figure~\ref{fig:fairness_regime} shows the results for enforcing \textsf{NaturalShare} and \textsf{PerSenderFairshare} policies with default parameter settings. We plot the ratio of clients' average window size to attackers' average window size  for the \textsf{NaturalShare} policy in Figure~\ref{fig:fairness_regime:a}.
For flat-rate attacks and shrew attacks, it may be surprising that the clients' average window size is larger than that of the attackers. Detailed trace analysis shows that it is because that the window sizes of a large portion of attackers keep decreasing, as we explained in our testbed experiment. As the number of attackers is much larger than the client count, the attackers' average window size turns out to be smaller than that of the clients, although the absolute volume of attack traffic may be still higher. Under reactive attacks, the clients' average window size (almost zero) is too small to be plotted in Figure \ref{fig:fairness_regime:a}.

Figure \ref{fig:fairness_regime:b} shows that the clients enjoy even larger window ratio gains under the \textsf{PerSenderFairshare} policy in flat-rate and shrew attacks because even more attackers enter the window dropping mode. Further, the \textsf{PerSenderFairshare} ensures that the clients' average window size is close to the per-sender fair rate in reactive attacks. Figure \ref{fig:fairness_regime:c} demonstrates such per-client fairness since Jain's fairness index~\cite{jain} (FI) is close to $1$.

\parab{\mb Coordination for Co-bottleneck Detection.} To enforce global per-sender fairness, the \mbx sharing the same bottleneck link share their local observations (\S \ref{sec:bandwidt_sharing_policy}). We first investigate how bad the FI can be without such inter-\mb coordination. We reconstruct the topology to create multiple \mbx, and map each client to a random \mb. The attackers launch reactive attacks. The results, plotted in Figure \ref{fig:fairness_regime:d}, show that the FI drops slightly, by ${\sim}8$\%, even if $20$ \mbx make local rate allocations without any coordination among them. 

\begin{figure}[t]
	\centering
	\mbox{
		\includegraphics[width=0.8\columnwidth]{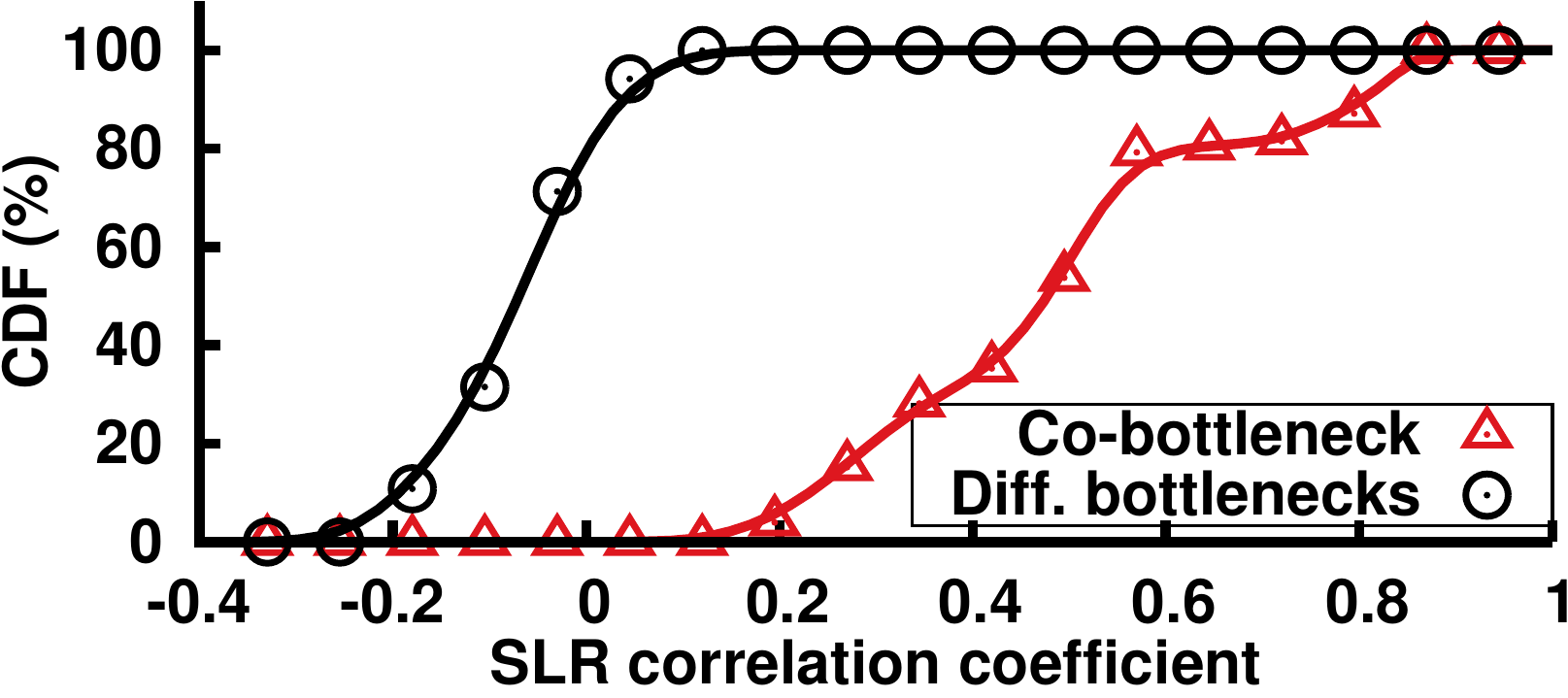}
	}
	\caption{[Simulation] The SLR correlation coefficient reflects whether two \mbx share a bottleneck link.}
	\label{fig:coeff}
\end{figure}

To complete our design, we further propose the following co-bottleneck detection mechanism. The design rationale is that if two \mbx' SLR observations are  correlated, they share a bottleneck with high probability. To validate this, we rebuild the network topology to create the  scenarios where two \mbx share and do not share a bottleneck, and study the correlation coefficient of their SLRs. We compute one coefficient for every 100 SLR measurements from each \mb. Figure~\ref{fig:coeff} shows the CDF of the coefficient. Clearly, the coefficient reflects whether  the two \mbx share a bottleneck. Thus, by continuously observing such correlation between two \mbx' SLRs, \sys can determine with increasing certainty whether or not they share a bottleneck, and can configure their coordination accordingly.

\begin{table}[t]
	\centering
	\subtable[The \textsf{NaturalShare} Policy]{
		\begin{tabular}{|c|c|c|c|c|c|c|}
			\hline
			\multirow{2}{*}{} & \multicolumn{2}{c|}{$\mathcal{D}_p$} & \multicolumn{2}{c|}{$\Th_{\slr}^{\drop}$} & \multicolumn{2}{c|}{$\beta$} \\ \cline{2-7} 
			& 2s         & 8s          & 0.03           & 0.1           & 0.5         & 0.9         \\ \hline
			Flat              & 1.1       & 0.17       & 0.78           & 0.39          & 1.1         & 0.78        \\ \hline
			Shrew             & 1.3       & 0.65       & 0.77           & 1.0           & 1.2         & 0.80        \\ \hline
		\end{tabular}
	}
	
	\subtable[The \textsf{PerSenderFairshare} Policy]{
		\begin{tabular}{|c|c|c|c|c|c|c|}
			\hline
			\multirow{2}{*}{} & \multicolumn{2}{c|}{$\mathcal{D}_p$} & \multicolumn{2}{c|}{$\Th_{\slr}^{\drop}$} & \multicolumn{2}{c|}{$\beta$} \\ \cline{2-7} 
			& 2s         & 8s          & 0.03           & 0.1           & 0.5         & 0.9         \\ \hline
			Flat              & 1.0      & 1.1       & 1.0           & 0.69          & 0.85         & 0.81      	\\ \hline
			Shrew             & 1.1       & 0.98       & 0.72           & 0.83           & 1.0         & 0.98        \\ \hline
			Reactive          & 1.0       & 0.99       & 1.0            & 0.94          & 1.0         & 1.0         \\ \hline
		\end{tabular}
	}
	\caption{[Simulation] Clients' average window size under different parameter settings.}\label{tab:para:dive}
\end{table}
 
\parab{Parameter Study.} In this section, we evaluate \sys using different parameters than the default values in Table \ref{tab:para}. 
We mainly focus on $\mathcal{D}_p$, $\Th_{\slr}^{\drop}$ and $\beta$. For each parameter, we vary its value in experiments to obtain the clients' average window size under the 10-million bot attack. The results showed in Table \ref{tab:para:dive} are normalized to the window sizes obtained using default parameters in Table \ref{tab:para}.

Under the \textsf{NaturalShare} policy, the shorter $\mathcal{D}_p$ produces a larger window size for legitimate clients since each sender's $\mathcal{W}_R$ is updated per-period so that a smaller $\mathcal{D}_p$ causes faster cut in attackers' window sizes. For $\Th_{\slr}^{\drop}$, a smaller value slows down the clients' recovery whereas a larger value allows larger  window sizes for attackers. Both will reduce the clients' share. A larger $\beta$ has negative effects as it takes more time for the clients to recover to a low LLR. 

With the \textsf{PerSenderFairshare} policy, \sys's performance is more consistent under different parameter settings. The most sensitive parameter is $\Th_{\slr}^{\drop}$ because it determines whether one source can send best-effort traffic.

%% file: related.tex
\begin{table*}[t]
	\centering
	\resizebox{\textwidth}{!}{
		\begin{tabular}{|c|c|c|c|c|c|c|c|c|c|}
			\hline
			\multirow{2}{*}{} 
			& Pushback\cite{pushback}                 
			& SIFF\cite{siff}, TVA\cite{TVA}
			& Netfence\cite{netfence}                                                                    
			& Phalanx\cite{phalanx}                                                          
			& Mirage\cite{mirage}                                                         
			& SIBRA\cite{SIBRA}
			& \sys
			\\ \hline
			Source upgrades   
			& \textbf{No}                          
			& Yes      
			& Yes                                                                         
			& Yes                                                                         
			& Yes 
			& Yes               
			& \textbf{No}                           
			\\ \hline
			Dest. upgrades     
			& \textbf{No}                                                                                                  
			& Yes                                                                         
			& Yes                                                                        
			& Yes                                                                           
			& Yes                                                         
			& Yes               
			& Yes                           
			\\ \hline
			AS deployment           
			& Remote and Unrelated          
			& Remote and Unrelated                                                                  
			& Remote and Unrelated                                                                   
			& Remote and Unrelated                                                                  
			& \textbf{Related}                                                                                                                  
			& Unrelated         
			& \textbf{Related}                       
			\\ \hline
			Router support    
			& $O(N)$ states      
			& \begin{tabular}[c]{@{}c@{}}Cryptography;\\$O(N)$ states for \cite{TVA}\end{tabular} 
			& \begin{tabular}[c]{@{}c@{}}$O(N)$ states;\\Cryptography\end{tabular} 
			& $O(N)$ states  
			& \begin{tabular}[c]{@{}c@{}}Larger\\memory\end{tabular}                                                      
			& \textbf{None}              
			& \textbf{None}                          
			\\ \hline
			\begin{tabular}[c]{@{}c@{}}Traffic control\\ policies\end{tabular}  
			& None               
			& None                                                                       
			& \begin{tabular}[c]{@{}c@{}}Per-sender\\ fairness\end{tabular}                                                               
			& None                                                                       
			&\begin{tabular}[c]{@{}c@{}}Per-compute\\fairness\end{tabular}                                                                                                           
			& \begin{tabular}[c]{@{}c@{}}Per-AS\\ fairness\end{tabular} 
			& \begin{tabular}[c]{@{}c@{}}\textbf{Victim-defined}\\ \textbf{policies}\end{tabular}                
			\\ \hline
			\begin{tabular}[c]{@{}c@{}}Other\\ requirements\end{tabular}         
			& \textbf{None}            
			& New header                     
			& \begin{tabular}[c]{@{}c@{}}New header;\\Passport\cite{passport}\end{tabular}                                                            
			& New header
			& \begin{tabular}[c]{@{}c@{}}Puzzle;\\ IPv6 upgrade\end{tabular}                                                       
			& \begin{tabular}[c]{@{}c@{}}Redesign\\ the Internet\end{tabular} 
			& \textbf{None}
			\\ \hline                      
		\end{tabular}
	}
	\caption{Property comparison with other research proposals. ``$O(N)$ 
		states'' means that the number of states maintained by a router  
		increases with the number of attackers. ``Cryptography'' means that 
		a router needs to support cryptography operation, \eg   
		MAC computation. ``Puzzle'' means that the mechanism requires 
		computational puzzle distribution. 
	}\label{tab:related}
	\normalsize
\end{table*}
\section{Related Work}\label{sec:related}
In this section, we briefly discuss previous academic work. Previous research approaches can be generally categorized into capability-based approaches (SIFF~\cite{siff}, TVA~\cite{TVA}, NetFence~\cite{netfence}), filtering-based approaches (Traceback~\cite{practicalIPTrace, advancedIPTrace}, AITF~\cite{AITF}, Pushback~\cite{pushback, implementPushback}, StopIt~\cite{StopIt}), overlay-based approaches (Phalanx~\cite{phalanx}, SOS~\cite{sos}, Mayday~\cite{mayday}), deployment-friendly approaches (Mirage~\cite{mirage}, CRAFT~\cite{craft}), approaches based on future Internet architectures (SCION~\cite{scion}, SIBRA~\cite{SIBRA}, XIA~\cite{xia}, AIP~\cite{aip}), and others (SpeakUp~\cite{speakup}, SDN-based~\cite{bohatei,sdns}, CDN-based~\cite{CDN_on_Demand}). We summarize the properties of one or two approaches from each category in Table~\ref{tab:related}. The comparison shows that \sys requires the \emph{least} deployment (no source upgrades, no additional router support and no deployment from unrelated ASes) while providing the \emph{strongest} property (enforcing destination-chosen traffic control policies). 

%% file: discussion.tex
\section{Discussion}\label{sec:discussion}
We briefly cover some aspects that are not previously discussed. 

\parab{Allowing ISPs to Disrupt the DDoS Prevention Industry.} A prerequisite of using existing DDoS prevention service providers is that a victim redirect its network traffic to these service providers. Cloudflare, for instance, will terminate all user SSL connections to the victim at Cloudflare's network edge, and then send back user requests (after applying their secret sauce filtering) to the destination server using new connections. Although this operation model is acceptable for small websites (\ie personal blogs), our industrial interview indicates that it is privacy invasive for large organizations such as government and hosting companies. Unless the victim fully trusts its selected cloud infrastructure or builds its own cloud infrastructure, \sys has the same privacy concern. Currently, these large organizations have to rely on their ISPs to block attack traffic. We leave it to future work on the exploration of how ISPs may disrupt the DDoS prevention industry.

\parab{\mbx Mapping.} \sys can leverage the end-user mapping \cite{end_user_mapping} to achieve better \mb assignment, such as redirecting clients to the nearest \mb, mapping clients according to their ASes, and load balancing.

\parab{Incorporating Endhost Defense.} \sys can cooperate with the DDoS defense mechanism deployed, if any, on the victim. For instance, via botnet identification \cite{bot_identify,bot_identify_2}, the victim can instruct the \mbx to block botnet traffic early at upstream so as to save more downstream bandwidth for clients. Such benefits are possible because the traffic control policies enforced by \sys are completely destination-driven. 

\parab{Additional Monetary Cost.} 
As discussed in \ref{sec:evaluation:overhead}, \sys introduces small computational overhead. Compared with basic DDoS-as-a-service solutions, \sys offers additional functionalities such as enabling destination-chosen policies and filtering bypassing traffic. In a competitive marketplace, service's price (the monetary cost) should scale with the cost of providing that service, which, in the case of \sys, is low. 

\parab{Traffic Learning Tools.} Through our interviews with industry people, we notice that some organizations have limited knowledge about their traffic. Thus, although \sys is able to enforce arbitrary victim-selectable traffic control policies, these organizations would receive limited benefits since they could not figure out correct policies (\eg what typical of traffic is most critical to their business, which clients are more important to their profit). To assist these organizations to better understand their network traffic and therefore bring out more rational policies, we are in active research on inventing multiple machine learning based traffic learning tools, focusing on traffic classification and client clustering.

%% file: conclusion.tex
\vspace*{-0.1in}
\section{Conclusion}\label{sec:conclusion}
Guided by our large-scale industrial interviews with potential DDoS victims, this paper presents \sys, the first DDoS mitigation system that offers readily deployable and proactive DDoS prevention. In its design, \sys explicitly addresses three challenges. First, \sys designs a capability mechanism that requires only limited deployment from the cloud, rather than widespread Internet upgrades. Second, \sys is fully destination-driven, addressing the shortcomings of the existing DDoS prevention systems that can only work either protocol or vendor defined traffic control policies. Finally, \sys addresses  the traffic-bypass vulnerability of the existing cloud-based solutions. Extensive evaluations on the Internet, testbed and large scale simulations validate \sys's deployability and effectiveness in enforcing detestation-chosen traffic control policies. Besides the technical contribution, based our large-scale survey, we also  articulated the design space of future advanced DDoS prevention mechanism, though which we hope to offer more insight into practical DDoS mitigation in research community.